\documentclass[prx,twocolumn,preprintnumbers,amsmath,amssymb,letterpaper,superscriptaddress]{revtex4}

\usepackage{graphicx}
\usepackage{latexsym}
\usepackage{amsmath}
\usepackage{graphics}
\usepackage{amssymb}
\usepackage{layout}
\usepackage{verbatim}
\usepackage{amsfonts,epsfig}
\usepackage{slashed}
\usepackage{feynmp}
\usepackage{float}
\usepackage{color}
\usepackage{hyperref}
\newcommand{\ket}[1]{\left|#1\right>}

\newcommand{\beq}{\begin{equation}}
\newcommand{\eeq}{\end{equation}}
\newcommand{\bea}{\begin{eqnarray}}
\newcommand{\eea}{\end{eqnarray}}
\newcommand{\nn}{\nonumber}

\newcommand{\mean}[1]{\langle{#1}\rangle{}}

\begin{document}

\title{Stabilization and manipulation of multispin states in quantum-dot time crystals with Heisenberg interactions} 

\author{Edwin Barnes}
\email{efbarnes@vt.edu}
\affiliation{Department of Physics, Virginia Tech, Blacksburg, Virginia 24061, USA}
\author{John M. Nichol}
\affiliation{Department of Physics and Astronomy, University of Rochester, Rochester, New York 14627, USA}
\author{Sophia E. Economou}
\affiliation{Department of Physics, Virginia Tech, Blacksburg, Virginia 24061, USA}

\begin{abstract}
A discrete time crystal is a recently discovered non-equilibrium phase of matter that has been shown to exist in disordered, periodically driven Ising spin chains. In this phase, if the system is initially prepared in one of a certain class of pure multi-spin product states, it periodically returns to this state over very long time scales despite the presence of interactions, disorder, and pulse imperfections. Here, we show that this phase occurs in GaAs quantum dot spin arrays containing as few as three quantum dots, each confining one electron spin, for naturally occurring levels of nuclear spin noise and charge noise. Although the physical interaction in these arrays is a nearest-neighbor Heisenberg exchange interaction, we show that this can be effectively converted into an Ising interaction by applying additional pulses during each drive period. We show that by changing the rotation axis of these pulses, we can select the direction of the Ising interaction and, consequently, the quantization axis of the stabilized multi-spin states. Moreover, we demonstrate that it is possible to perform coherent rotations of the stabilized states while remaining in the time crystal phase. These findings open up the intriguing possibility of using time crystal phases to extend the lifetime of quantum states for information applications.
\end{abstract}

\maketitle

\section{Introduction}

\begin{figure}
\includegraphics[width=0.9\columnwidth]{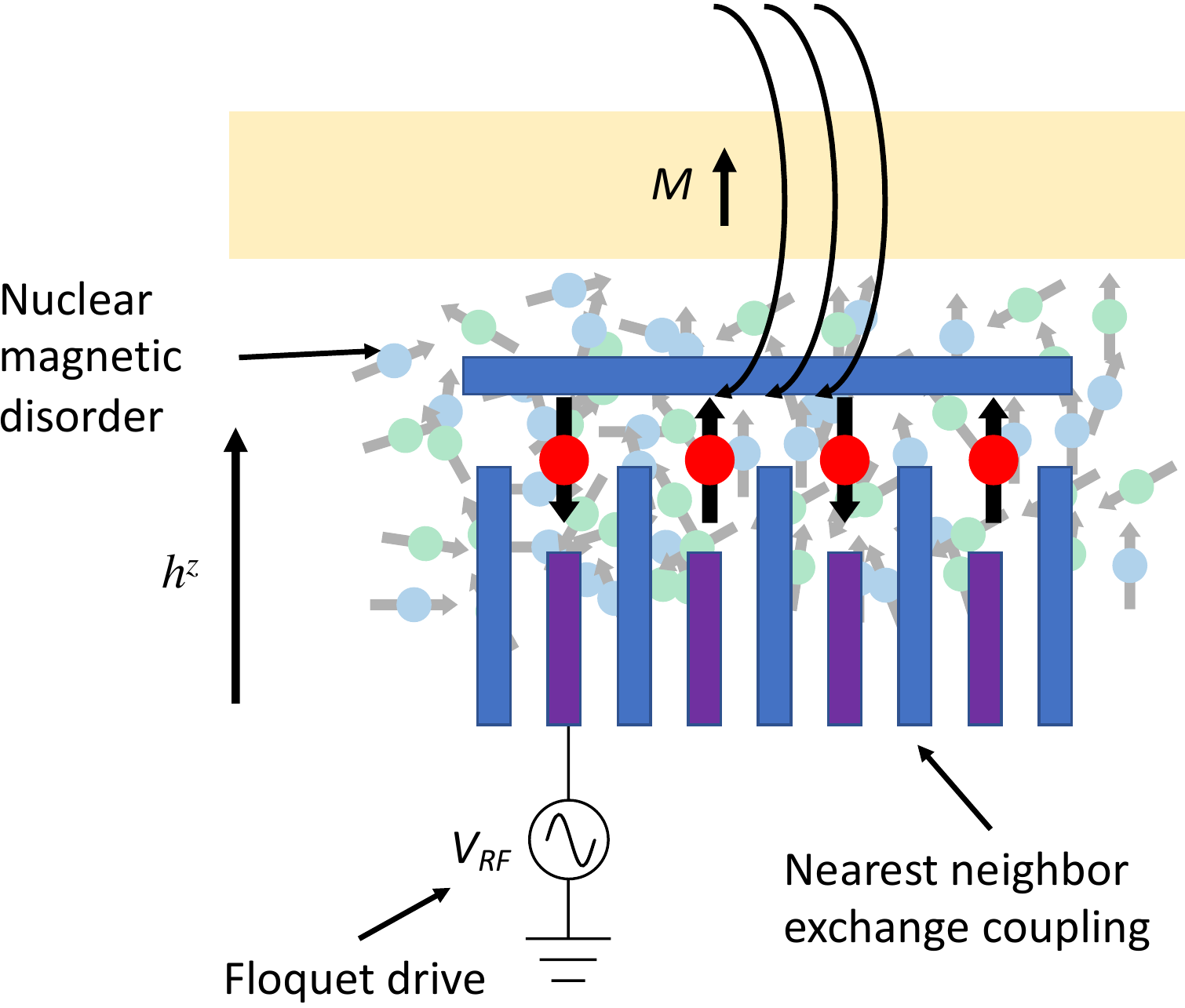}
\caption{$N=4$ GaAs quantum dot array. Each dot contains one electron spin that couples to a bath of nuclear spins in the surrounding semiconductor. An external magnetic field is applied along the $z$ direction to split the spin states. A micromagnet (yellow) generates a magnetic field gradient $dh^x/dz$ across the chain. A voltage modulation $V_{RF}$ induces displacement of the electron in the $z$ direction and an oscillating transverse magnetic field, causing spin rotations.}\label{fig:cartoon}
\end{figure}

The concept of spontaneous symmetry breaking---the phenomenon in which the ground state of a system does not preserve symmetries of the Hamiltonian---lies at the heart of many classic phases of matter such as crystalline solids, ferromagnetism, and superconductivity \cite{symmetrybreaking}. These phases break a variety of symmetries, including invariance under rotations, spatial translations, or gauge transformations. In 2012, it was pointed out by Wilczek \cite{Wilczek_PRL12} that a notable omission from this list is time translation invariance, and he went on to propose a class of systems which could potentially realize continuous time-translation symmetry breaking \cite{Li_PRL12,Wilczek_PRL13}. Although a series of no-go theorems \cite{Bruno_PRL13a,Bruno_PRL13b,Watanabe_PRL15} later led to the conclusion that continuous time-translation symmetry breaking is not possible, this in turn motivated the question of whether it is possible to have systems which exhibit {\it discrete} time-translation symmetry breaking in which periodically driven systems are characterized by observables that do not evolve with the same periodicity as the Hamiltonian. Recent seminal works \cite{Else_PRL16,Khemani_PRL16} demonstrated that this is indeed possible provided one defines such phases, known as discrete time crystals, appropriately. This is a subtle issue because, unlike other types of spontaneous symmetry breaking, discrete time crystals cannot be defined with respect to a ground state since energy is not conserved. Else et al. proposed \cite{Else_PRL16} to define a discrete time crystal as a phase in which there exists an observable for which all spatially short-range correlated states break the time periodicity of the Hamiltonian. They further showed that a periodically driven Ising spin chain with local disorder satisfies this definition. In this example, the presence of disorder is important for bringing the system into a many-body localized phase \cite{Basko_AoP06,Pal_PRB10,Nandkishore_ARCMP15}, which guarantees that multi-spin states remain non-ergodic and short-range correlated, in turn preventing the system from heating up into a thermal state despite the presence of interactions and driving. Soon after these theoretical proposals, experimental demonstrations of discrete time crystal phases were performed in several different physical systems containing ten or more pseudo-spin 1/2 particles with Ising interactions, including trapped ions \cite{Zhang_Nature17}, NV centers in diamond \cite{Choi_Nature17}, and P donors in silicon \cite{Xu_PRL18}.

Gate-defined quantum dot arrays (Fig.~\ref{fig:cartoon}) provide a natural platform in which to realize time crystals owing to recent progress in fabrication and control of larger arrays consisting of four or more dots, each containing a single electron spin that is exchange coupled with its nearest neighbors \cite{Ito_arxiv18,Ito_SciRep16,Mukhopadhyay_APL18,Hensgens_Nature17,Mills_arxiv18}. Moreover, in the case of GaAs quantum dots, such systems naturally feature on-site magnetic field disorder coming from the three different species of spinful nuclei in GaAs \cite{Cywinski_PRB09,Bluhm_NP11,Malinowski_NatNano17}. There are on the order of $10^6$ nuclear spins in each quantum dot, and the statistical polarization distribution of these spins provides an effective, local magnetic field (known as the Overhauser field \cite{Overhauser_PR53}) that acts on the electronic spins via contact hyperfine interactions. The presence of charge noise in these systems also leads to random variations in the spin-spin couplings along the chain, which can provide some additional stability to time crystal phases \cite{Else_PRL16,Yao_PRL17}. The natural levels of nuclear spin noise and charge noise are already sufficiently strong to bring such arrays into the many-body localized phase underlying the time crystal, as shown in a recent work by one of the authors \cite{Barnes_PRB16}, suggesting that a time crystal could be experimentally realized by applying periodic driving.

In this paper, we show with numerical simulations that discrete time crystal phases can be realized in small GaAs quantum dot spin arrays containing as few as three quantum dots with one electron trapped in each. Although the natural interactions in quantum dot spin arrays are nearest-neighbor Heisenberg exchange interactions, we show that these can be effectively converted into Ising interactions by including additional pulses in each period of the drive. We compute non-equilibrium ``phase diagrams" to determine the range of interaction strengths and pulse errors over which robust time crystal phases are realized. We demonstrate that by choosing the rotation axes of these pulses, we can select which multi-spin states are preserved by the time crystal. {\it Thus, Heisenberg interactions allow for the preservation of a much larger class of multi-spin states compared to the Ising case.} Furthermore, we show that we can rotate between different multi-spin states by modifying the driving sequences without disturbing the time crystal phase. These results suggest that it may be possible to use time crystal phases to stabilize and manipulate quantum states for information applications. Although our focus is on quantum dot arrays, our results are applicable for any system that is well described by a disordered Heisenberg spin chain.

The paper is organized as follows. In Sec.~\ref{sec:ising}, we review the Ising model of Ref.~\cite{Else_PRL16} and re-analyze the time crystal phase for the case of small spin chains containing four spins. In this context, we introduce our methods for distinguishing phases in the few-spin case and study how these phases depend on various system and noise parameters. This sets the stage for a direct comparison to the Heisenberg spin chain case investigated in Sec.~\ref{sec:heisenberg} and relevant to quantum dot spin arrays. There, we show how to recover effective Ising interactions using additional pulses, which we refer to as ``H2I pulses", during each drive period and compare the resulting phase diagrams to the pure Ising case. We also examine to what extent different initial states are preserved by the time crystal phase. In Sec.~\ref{sec:control}, we show how switching the axis of the additional pulses appropriately can generate coherent spin rotations without disrupting the time crystal phase. We also investigate the preservation of quantum superpositions. We conclude in Sec.~\ref{sec:conclusions}.

\section{Time crystals in Ising spin chains}\label{sec:ising}

Else et al. \cite{Else_PRL16} defined a discrete time crystal as a phase of a periodically driven system, with Hamiltonian $H(t+T)=H(t)$ and drive period $T$, in which there exists some observable $\cal O$ for which all spatially short-range correlated states $\psi$ break the periodicity of the Hamiltonian: $\mean{{\cal O}(t+T)}_\psi\ne\mean{{\cal O}(t)}_\psi$. Since the expectation value of any operator taken with respect to an eigenstate of the Floquet operator, ${\cal U}(T)={\cal T}e^{-i\int_0^TdtH(t)}$, will necessarily be periodic with period $T$, this definition implies that all Floquet eigenstates must be spatially long-range correlated in the time crystal phase. They further showed that the following disordered, driven Ising spin chain,
\begin{eqnarray}
H_{ITC}(t)&=&\sum_{i=1}^N(J_i\sigma_i^z\sigma_{i+1}^z+h_i^z\sigma_i^z)\nonumber\\&&-(\pi/2-\epsilon)\sum_{k=1}^\infty\delta(t-kT)\sum_{i=1}^N\sigma_i^x,\label{isingham}
\end{eqnarray}
satisfies this criterion, yielding an Ising time crystal. Here, $\sigma_i^\alpha$ are Pauli operators for the $i$th spin, and the coupling constants $J_i$ and local fields $h_i^z$ are random variables distributed uniformly over the intervals $J_i\in[J-\delta J,J+\delta J]$, $h_i^z\in[h^z-\delta h^z,h^z+\delta h^z]$, for some choice of constant parameters $J$, $h^z$. The random fields $h_i^z$ are included to show that the time crystal phase is robust against moderate levels of local disorder. Time crystals can also be realized for transverse random fields, in which case the undriven system can exhibit a many-body localized phase \cite{Pal_PRB10,Luitz_PRB15,Nandkishore_ARCMP15}, which can help prevent the system from thermalizing despite the continual driving \cite{Else_PRL16}. This will become important in the next section, where we consider various orientations of the local disorder and effective Ising interaction. The second line of Eq.~\eqref{isingham} contains periodic $\pi$ pulses that break continuous time-translation symmetry down to discrete translations by $T$. These pulses, which we will refer to as Floquet pulses since they define the Floquet period, are reminiscent of dynamical decoupling pulses \cite{Hahn_PR50,Carr_Purcell,Meiboom_Gill,Viola_PRA98} and indeed counteract the longitudinal disorder $h_i^z$ and refocus the spins after every second period. However, as shown in Ref.~\cite{Yao_PRL17}, this refocusing effect is not robust against rotation errors $\epsilon$ unless the nearest-neighbor interactions $J_i$ are sufficiently strong. This is a striking signature that has been used to experimentally confirm the existence of time crystal phases in trapped ion systems \cite{Smith_NatPhys16}. Ref.~\cite{Yao_PRL17} used this phenomenon to map out a phase diagram for the system as a function of $J$ and $\epsilon$, where they found that for moderate values of $\epsilon$ on the order of 0.1, the system transitions from a symmetry unbroken MBL phase at small $J$ to a time crystal phase at moderate $J$, and then to a thermal phase at large $J$.

\begin{figure}
\includegraphics[width=0.49\columnwidth]{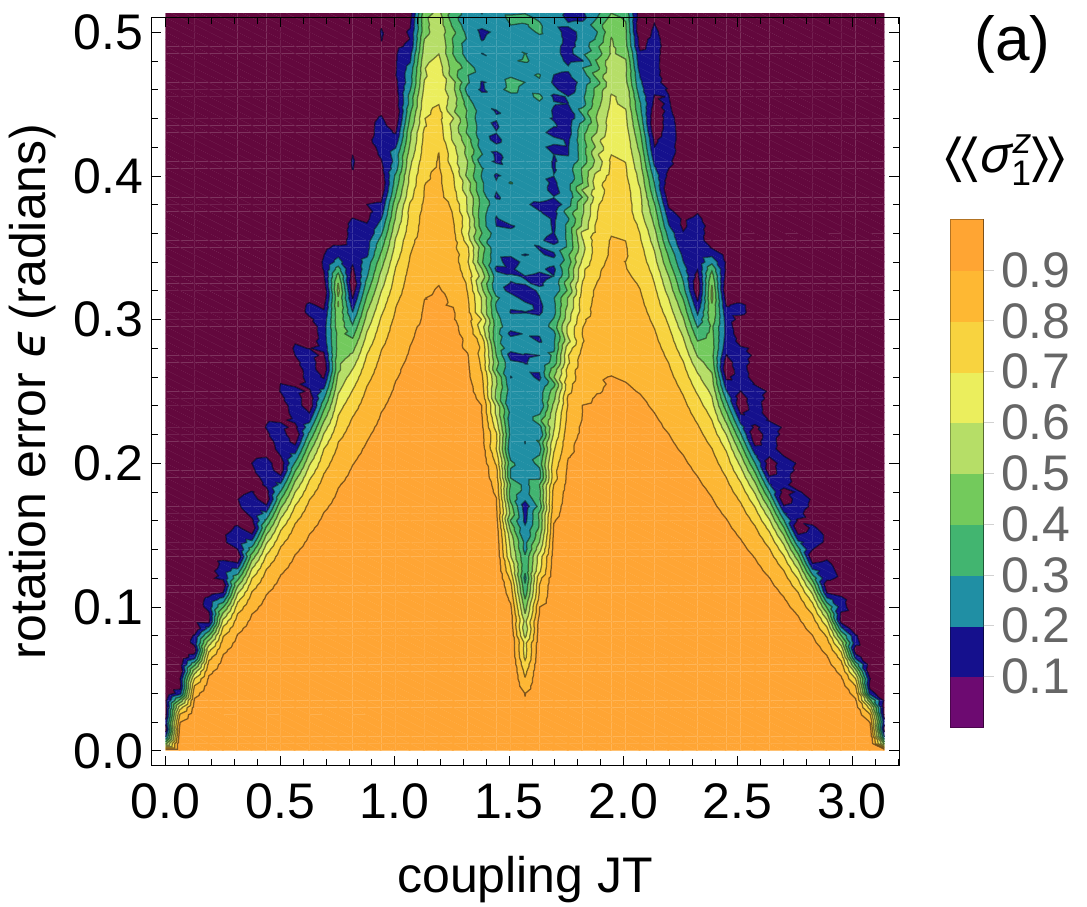}
\includegraphics[width=0.49\columnwidth]{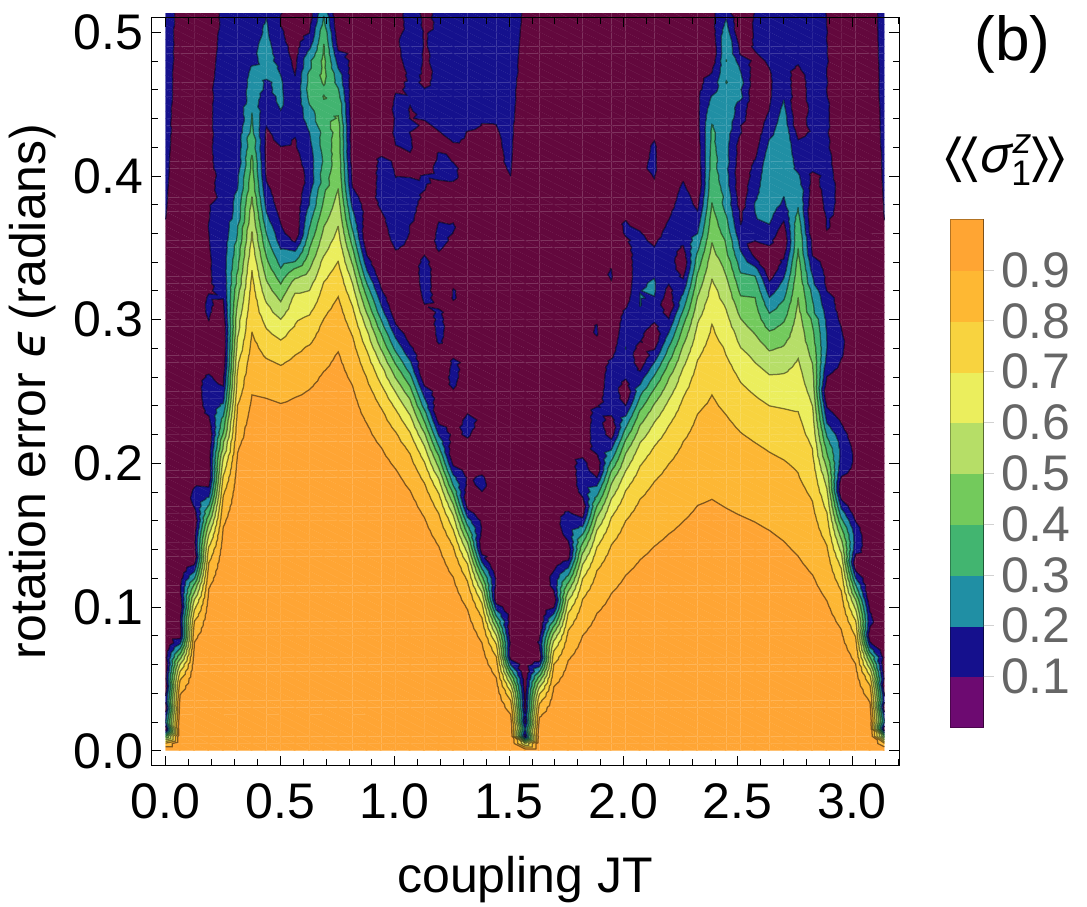}
\includegraphics[width=0.49\columnwidth]{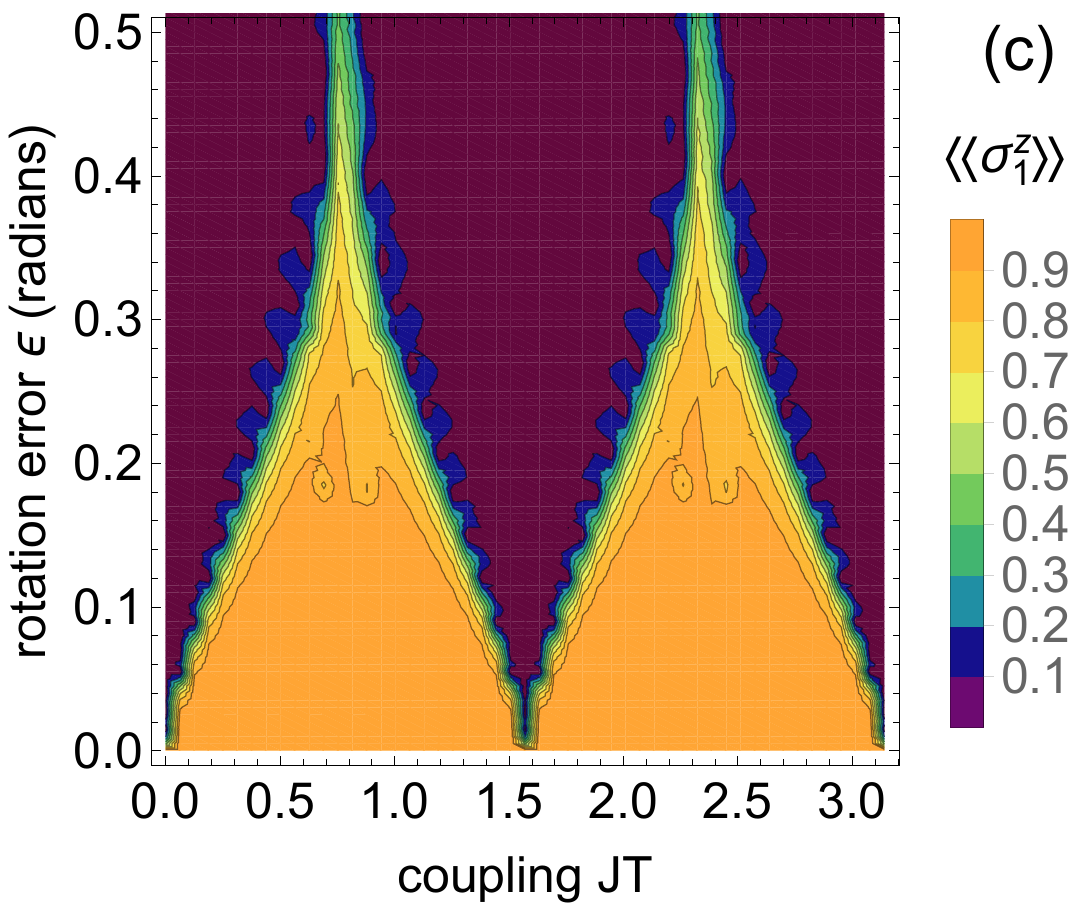}
\caption{(a), (b) The time-averaged $z$-component of the spin operator for the left-most spin of an open $N=4$ Ising spin chain initialized in the state $\ket{\uparrow\downarrow\uparrow\downarrow}$ as a function of coupling strength $J$ and rotation error $\epsilon$. The parameters are $\delta JT=0$, $\delta h^zT=0.05$ and (a) $h^zT=0.05$, (b) $h^zT=1$.  The orange regions correspond to time crystal phases in which the system returns to its initial state periodically with period $2T$. (c) The time-averaged $z$-component of the spin operator of a spin in a $2\times2$ Ising-coupled square quantum dot array initialized in the state $\ket{\uparrow\downarrow\uparrow\downarrow}$ with $h^zT=0.05$, $\delta h^zT=0.05$, $\delta JT=0$.}\label{fig:isingpd1}
\end{figure}

\begin{figure*}
\includegraphics[width=0.6\columnwidth]{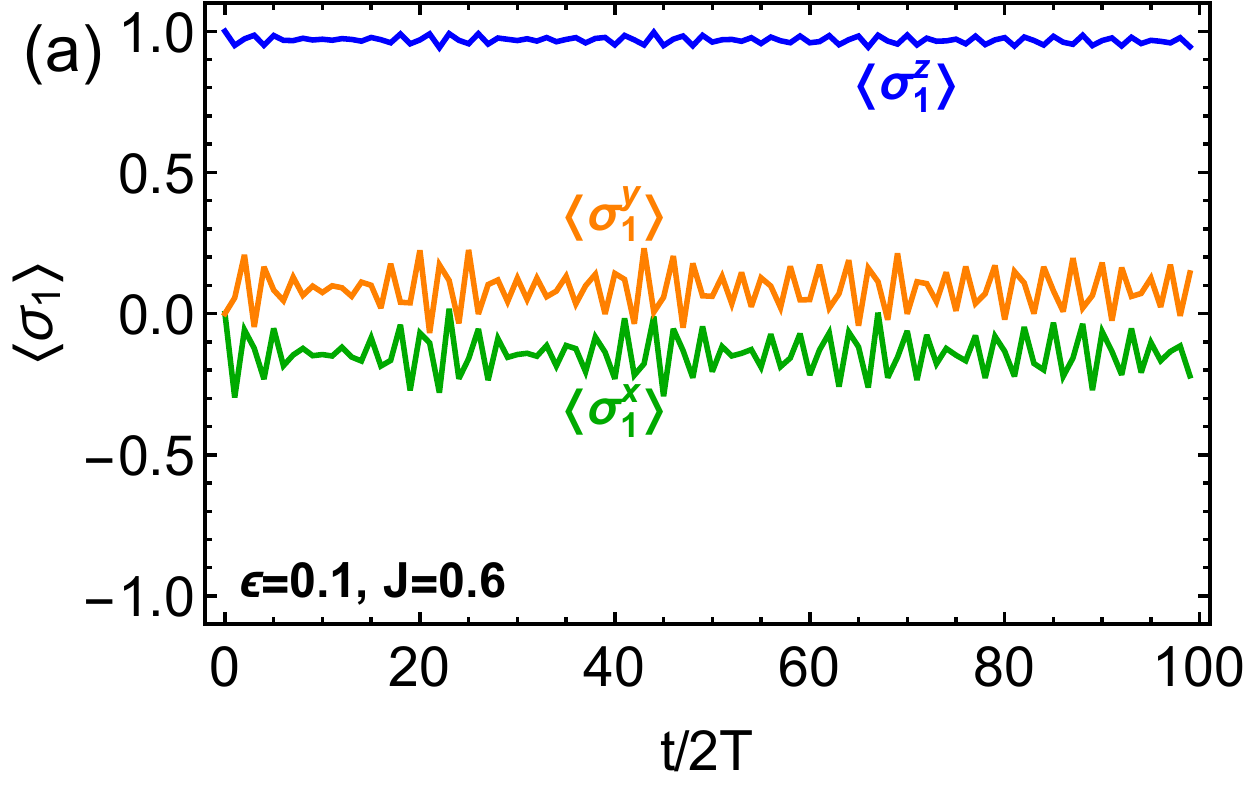}
\includegraphics[width=0.6\columnwidth]{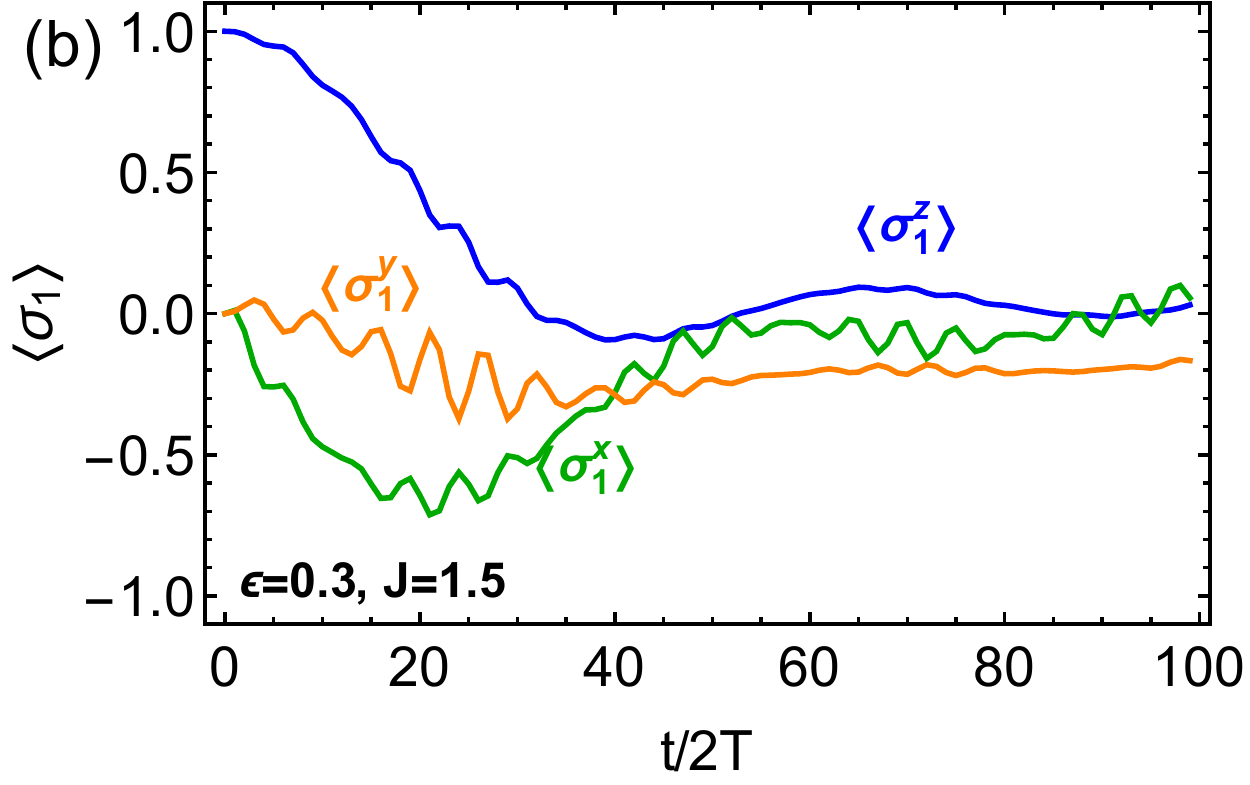}
\includegraphics[width=0.6\columnwidth]{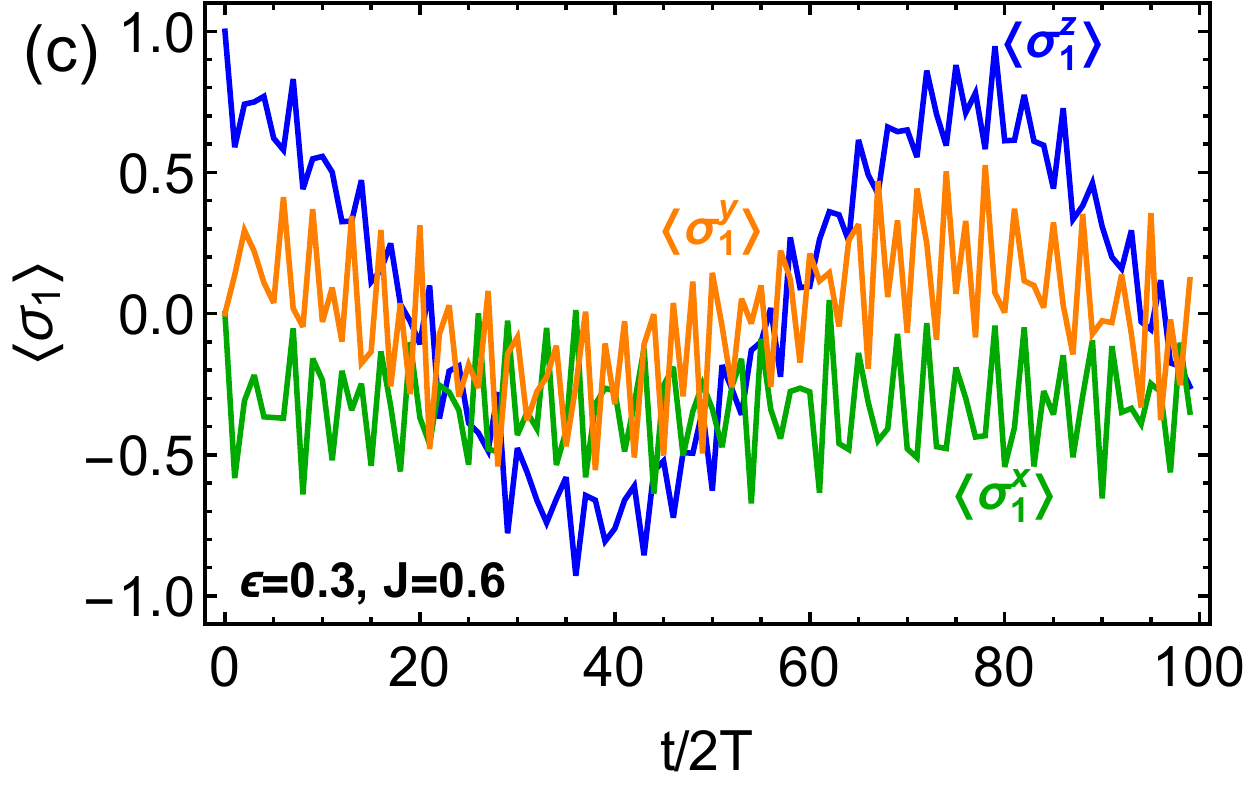}
\caption{Time evolution of spin vector components of first spin in $N=4$ Ising chain from Eq.~\eqref{isingham} in (a) time crystal phase, (b) thermal phase, (c) MBL phase. The initial state is $\ket{\uparrow\downarrow\uparrow\downarrow}$. The parameters are as in Fig.~\ref{fig:isingpd1}(a): $h^zT=0.05$, $\delta h^zT=0.05$, and $\delta JT=0$. The spin vector is shown only at discrete times $t_m=2mT$.}\label{fig:pointplots_ising}
\end{figure*}

In the present work, our focus is on the state-preservation properties afforded by the time crystal phase. Specifically, if the system is initially prepared in a product state in which each spin is in either $\ket{\uparrow}$ or $\ket{\downarrow}$ in the $z$-basis, then the system returns to this state after every second period, i.e., at times $t=2mT$ for all positive integers $m$. We can track whether or not this preservation occurs for a given set of system and control parameters by checking whether the time average of the $z$-component of one spin---say the first spin located at one end of an open spin chain---is close to unity at late times: $\mean{\mean{\sigma_1^z}}=\tfrac{1}{\ell+1}\sum_{m=0}^\ell\mean{\sigma_1^z(2mT)}\approx1$, where $\ell\gg1$. Throughout this paper, we take $\ell=100$, corresponding to 200 Floquet periods. Calculating this average for a range of system parameters provides us with an effective ``phase diagram". Note that this phase diagram is distinct from the one considered in Ref.~\cite{Yao_PRL17}, where a series of alternative diagnostics were used to identify the phase boundaries. We can of course also consider tracking other spins in the chain; this yields qualitatively similar phase diagrams as described below.

In Fig.~\ref{fig:isingpd1}, we show such a phase diagram for the case of an $N=4$ Ising chain in the absence of coupling disorder ($\delta J=0$). Panels (a) and (b) show the result for an open spin chain for two different values of the external magnetic field. (We define our units in terms of $T$ throughout this paper. The issue of units will be revisited in the next section when we consider realistic noise levels in exchange-coupled quantum dot arrays.) We see from the diagrams that regardless of the strength of the applied field, there is a large region (shaded orange) of parameter space in which the state of the spin is preserved, i.e., $\mean{\mean{\sigma_1^z}}\approx1$. We can identify this region as a time crystal phase, as is consistent with Ref.~\cite{Yao_PRL17}. If we look at the time evolution of the spin vector for the end spin in this region, we find that it remains essentially constant, as shown in Fig.~\ref{fig:pointplots_ising}(a). In this figure, we only show the spin vector values after every second Floquet period; to confirm that the $T$-translation symmetry is indeed broken, we show the full, continuous time evolution in Appendix~\ref{app:DTTSB}.

\begin{figure}
\includegraphics[width=0.6\columnwidth]{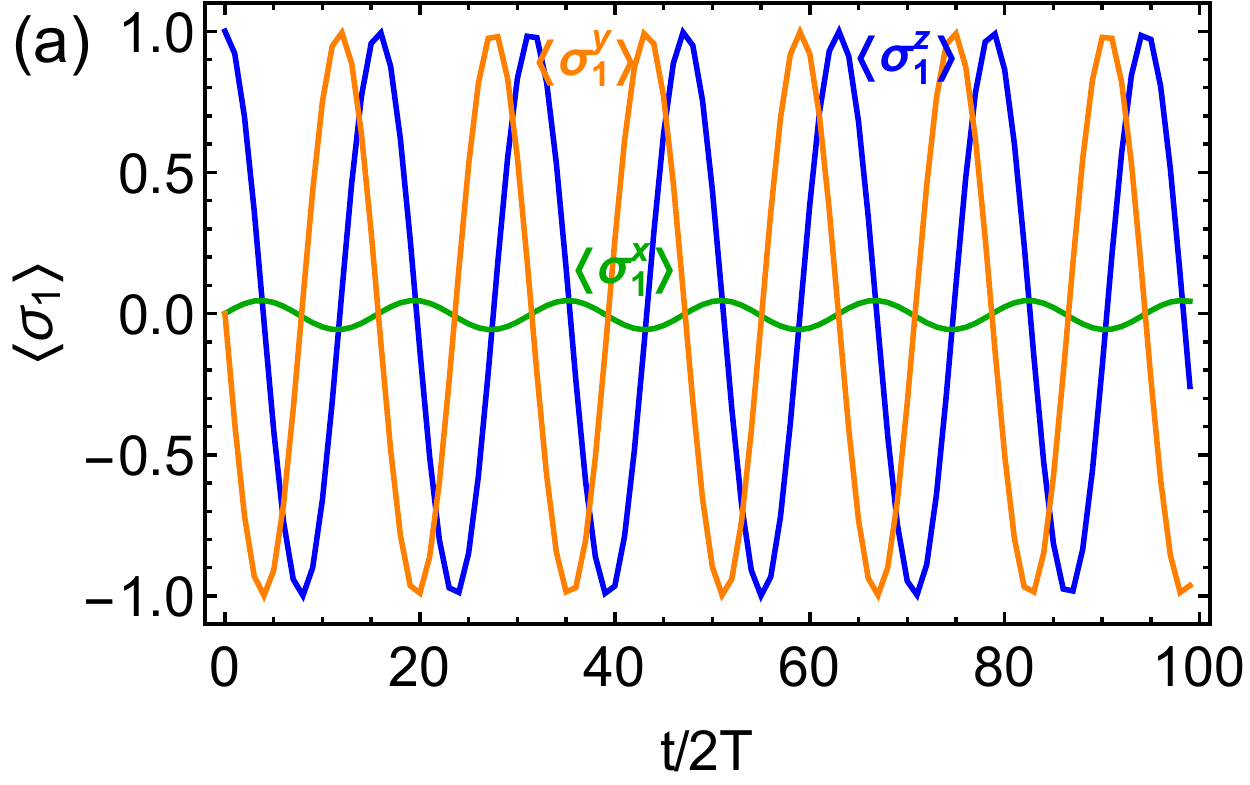}
\includegraphics[width=0.6\columnwidth]{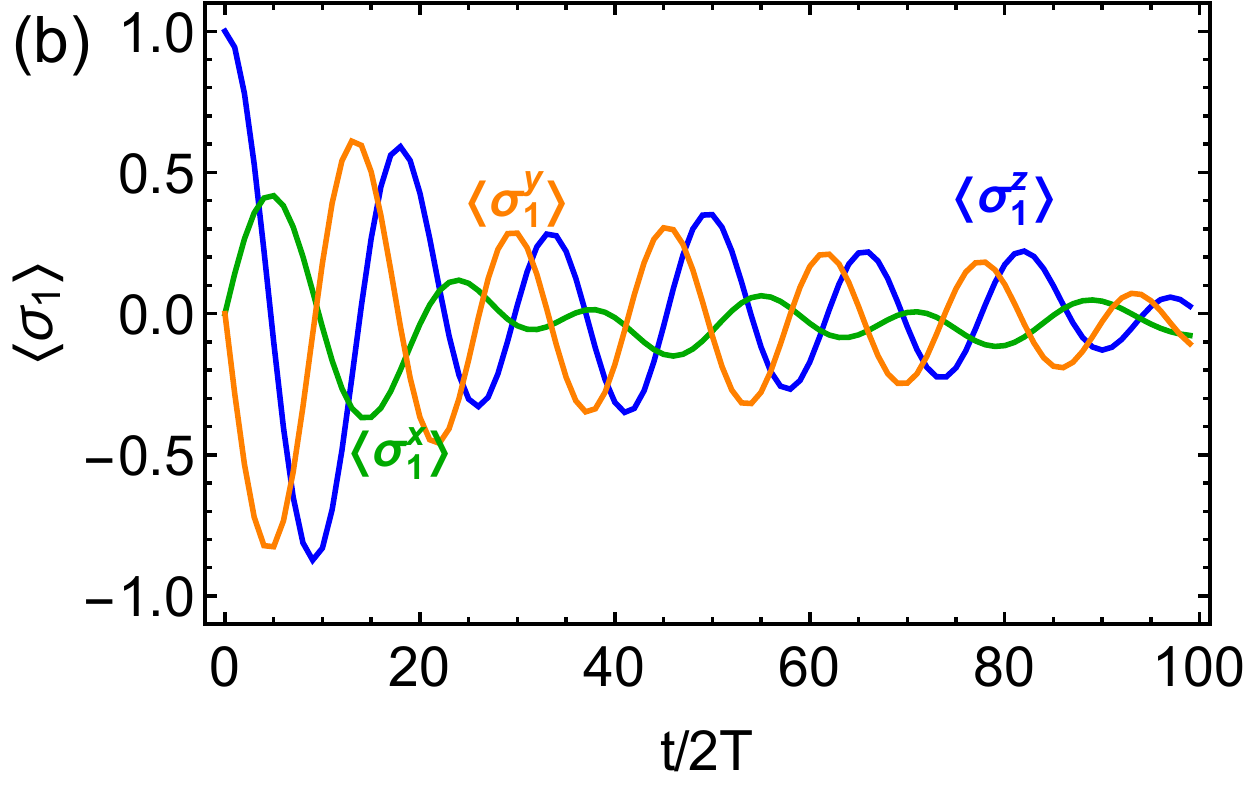}
\caption{Time evolution of spin vector components of first spin in $N=4$ Ising chain from Eq.~\eqref{isingham} with zero coupling ($J=0$, $\delta J=0$). The initial state is $\ket{\uparrow\downarrow\uparrow\downarrow}$. The Floquet pulse has error $\epsilon=0.1$, and the local field parameters are (a) $h^zT=0.05$, $\delta h^zT=0.05$, (b) $h^zT=0.5$, $\delta h^zT=0.5$. The spin vector is shown only at discrete times $t_m=2mT$.}\label{fig:pointplots_ising_nocoupling}
\end{figure}

There are also large regions near $J=0$ and $J=\pi$ (shaded brown) where the average spin component is close to zero. The region at small values of $J$ can be identified with the MBL phase found in Ref.~\cite{Yao_PRL17}. This region repeats at $J=\pi$ since the Floquet operator remains unchanged when $J$ is shifted by an integer multiple of $\pi$ (provided there is no coupling disorder, $\delta J=0$). This periodicity only occurs in the case of Ising interactions---it does not hold for Heisenberg interactions as we will see. The blue peninsula in Fig.~\ref{fig:isingpd1}(a) can be interpreted as a thermal phase. The fact that $\mean{\mean{\sigma_1^z}}$ assumes a low but nonzero value in this region is because $\mean{\sigma_1^z(t)}$ quickly decays to zero, as shown in Fig.~\ref{fig:pointplots_ising}(b). This is qualitatively different from the behavior of this quantity in the MBL phase, where it instead exhibits slow oscillations with an amplitude that remains large at late times (Fig.~\ref{fig:pointplots_ising}(c)). In this phase, the spins remain coherent as in Ref.~\cite{Barnes_PRB16}, but the pulses rotate the spins and break the $2T$-periodicity exemplified in the time crystal phase. The disappearance of the thermal phase around $J=\pi/2$ in Fig.~\ref{fig:isingpd1}(b) can be attributed to the fact that as the external field exceeds the coupling strength, the interactions become ineffective at mixing different non-interacting (product) eigenstates. We have checked that as the number of spins is increased, the crossover regions between the different phases in Fig.~\ref{fig:isingpd1}(a) become sharper, which is consistent with the findings of Ref.~\cite{Yao_PRL17}.

In Fig.~\ref{fig:isingpd1}(c), we show the phase diagram for a two-dimensional $N=4$ square quantum dot array rather than a linear array as we have been considering thus far. While the time crystal region remains similar in size to the linear array case, two differences arise for the square array. The first is that the phase diagram becomes symmetric about $J=\pi/2$, and the time crystal regions become more triangular. The second difference is that the thermal phase near $J=\pi/2$ no longer appears and is instead replaced by an additional MBL region that extends all the way down to $\epsilon=0$. Both differences are due to the fact that when the spin chain forms a closed loop, the Floquet  operator now becomes invariant under $\pi/2$ shifts of the coupling $J$ instead of the $\pi$-shift invariance we have in the case of an open chain. We find a similar phase diagram if we instead consider the spin vector component of an inner spin rather than the end spin. Thus the asymmetry about $J=\pi/2$ evident in Figs.~\ref{fig:isingpd1}(a),(b) is a consequence of the boundary of the open chain. For both open and closed spin chains, we have also considered different initial states such as $\ket{\uparrow\uparrow\downarrow\downarrow}$ and $\ket{\uparrow\uparrow\uparrow\uparrow}$ as well as different numbers of spins ranging from $N=3$ to $N=6$, but find only minor quantitative deviations from the phase diagrams in Fig.~\ref{fig:isingpd1} in all cases. A more significant dependence on the initial state and on $N$ arises in the context of Heisenberg spin chains, as is discussed below.

The role of the spin-spin coupling $J$ can be understood as counteracting the Floquet pulse error $\epsilon$ \cite{Choi_Nature17}. This is evident if we first tune the system into the time crystal phase and then switch off the coupling. Two examples of the resulting evolution are shown in Fig.~\ref{fig:pointplots_ising_nocoupling}, one for weak ($h^z=0.05$, $\delta h^z=0.05$) and the other for strong ($h^z=0.5$, $\delta h^z=0.5$) local longitudinal field disorder. In the former case, although the Floquet pulse has a nonzero error, it still cancels the longitudinal field noise and prevents dephasing. Thus, the spins remain coherent, and the only effect of nonzero $\epsilon$ is to evolve the spins away from their initial state and break the $2T$-periodicity. As shown in Fig.~\ref{fig:pointplots_ising}(a), when the coupling is switched on, the system returns to  its initial state with period $2T$ as if the pulse error has been effectively removed. When the coupling is turned off in the strong disorder case, the Floquet pulses are no longer capable of preventing dephasing due to the nonzero $\epsilon$ as shown in Fig.~\ref{fig:pointplots_ising_nocoupling}(b). Switching the couplings back on leads to a result similar to that shown in Fig.~\ref{fig:pointplots_ising}(a) and again causes the system to act as though there were no error in the Floquet pulses.

\section{Time crystals in Heisenberg quantum dot spin chains}\label{sec:heisenberg}

A semiconductor quantum dot spin array is well described by the following Heisenberg spin chain model \cite{Hensgens_Nature17}:
\begin{equation}
H_{Heisenberg}=\sum_{i=1}^N\left(J_i\sum_\alpha\sigma_i^\alpha\sigma_{i+1}^\alpha+h_i^z\sigma_i^z\right),\label{heisenbergham}
\end{equation}
where the $J_i\in[J-\delta J,J+\delta J]$ are now interpreted as inter-dot exchange couplings induced by quantum dot wave function overlaps (i.e. tunnel couplings between dots), and $h_i^z\in[h^z-\delta h^z,h^z+\delta h^z]$ is a local magnetic field that varies from dot to dot. Disorder in the $J_i$ generically arises from charge noise as has been seen experimentally in a variety of host materials \cite{Dial_PRL13,Martins_PRL16,Reed_PRL16,Nichol_npjQI17,Petit_arxiv18,Hu_PRL06}, while in the case of GaAs quantum dots, the local magnetic field disorder arises naturally from hyperfine interactions between the electronic spins and the nuclear spins intrinsic to the material \cite{Medford_PRL12,Fink_PRL13}. Although the fluctuations in the nuclear spin bath would generally cause effective local magnetic field variations in all directions (and not just $z$), these fluctuations are sufficiently slow that the electron spins follow these variations as they would an adiabatically evolving magnetic field; we can then redefine the direction of the total magnetic field (applied field plus Overhauser field) to be $z$, and the main effect of nuclear spin fluctuations is to randomly change the magnitude of the total field \cite{Bluhm_NP11,Neder_PRB11,Malinowski_NatNano17}. It was shown in Ref.~\cite{Barnes_PRB16} that the levels of both charge and nuclear spin noise in GaAs quantum dot arrays are sufficient to bring the system into a many-body localized phase for typical values of exchange couplings. This suggests that a time crystal phase can also be realized in such systems. Here, we will show that this is indeed the case provided the system is driven in a specific way.

Although both nuclear spin noise and charge noise in GaAs quantum dots have been measured to have a nontrivial power spectrum, it has been shown that they can be well described in terms of quasistatic noise \cite{Martins_PRL16,Barnes_PRB16b}. In this approximation, the effect of the noise is incorporated as stochastic shifts of Hamiltonian parameters such as $h_i^z$ and $J_i$ that are held constant over a single run of an experiment, but which are allowed to vary according to some distribution over the course of many runs. Decoherence effects then arise when the results of these runs are averaged. This way of modeling noise is identical to how local disorder is typically incorporated in spin-chain-based studies of time crystals, as reviewed in the previous section. Thus, although our focus is on quantum dot arrays, the analysis we present here is in fact quite general and could be applied to any physical system that is well described by a disordered Heisenberg spin chain.

Using nuclear spin programming and Bayesian estimation, it has been shown in double quantum dots that the width of the distribution for nuclear spin noise, $\delta h^z$, can be reduced from a few tens of MHz down to a few tens of kHz \cite{Foletti_NP09,Bluhm_PRL10,Shulman_NatCommun14}. The width of the distribution for charge noise, $\delta J$, can also be reduced by raising and lowering the inter-dot barrier rather than tilting the double well potential in order to control the exchange interaction \cite{Martins_PRL16,Reed_PRL16,Bertrand_PRL15}. Here, we focus on relatively high levels of nuclear spin noise (tens of MHz) and low levels of charge noise (fluctuations in $J$ below the one percent level) because we expect that this regime is the most relevant for quantum dot arrays, where barrier-controlled exchange couplings will be needed to achieve sufficient control over the system, and nuclear spin programming will be challenging to implement. We will continue to define our units with respect to the drive period, $T$. Since we are treating the pulses as $\delta$-functions, we are implicitly assuming that the duration of each pulse is much smaller than $T$. Given that pulse durations are on the order of 10 ns or less in current experiments \cite{Yoneda_PRL14}, this implies that the drive period should be on the order of 1 $\mu$s or more. Choosing $T=1$ $\mu$s then implies that energies are measured in units of $2\pi$ MHz. Thus, we focus on values of $\delta h^z$ on the order of 50-100 and values of $\delta J$ on the order of $0.01 J$ or less. We can also estimate typical values of the pulse rotation angle error $\epsilon$. If we assume that this error is due primarily to fluctuations in the electron spin splitting (and hence in the drive detuning) caused by $\delta h^z$, then this error is related to the dephasing time, $T_2^*$, according to $\epsilon\approx\tfrac{2\ln2}{\pi}\tau^2/T_2^{*2}$, where $\tau$ is the pulse duration. Assuming a $\pi$-pulse duration of $\tau=4$ ns \cite{Yoneda_PRL14} and a dephasing time of $T_2^*=10$ ns, this yields a pulse error of $\epsilon\approx0.07$.

\begin{figure}
\includegraphics[width=\columnwidth]{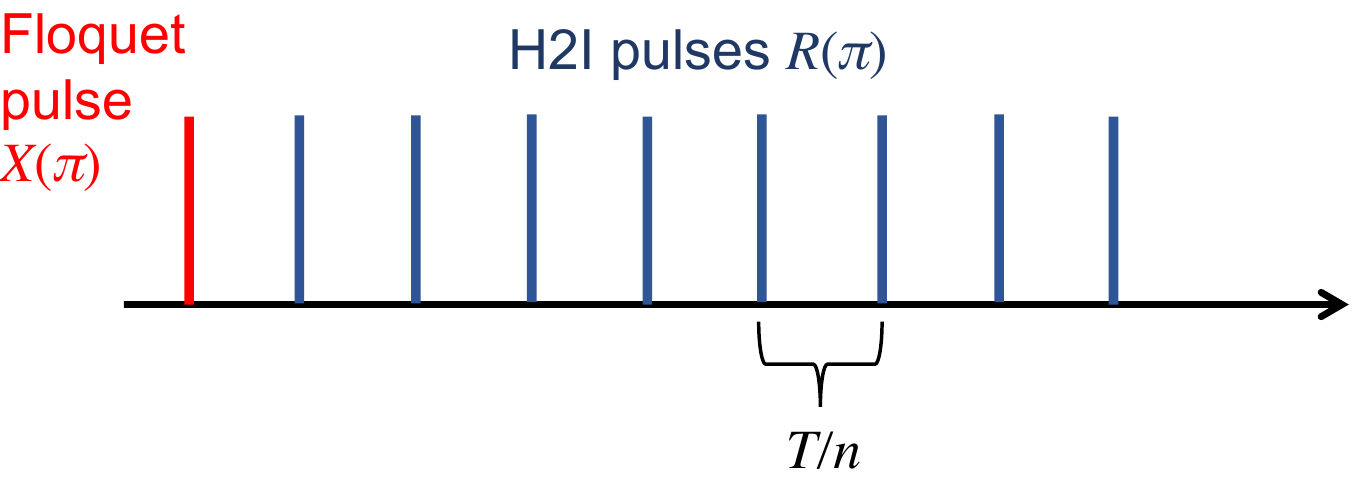}
\caption{Pulses applied during a single Floquet period of duration $T$, including one Floquet $\pi$-pulse about $x$ and $n$ H2I $\pi$-pulses. Here, the case $n=8$ is depicted. The axis of the H2I pulses can be chosen at will, and it determines the orientation of the effective Ising interaction.}\label{fig:H2Icartoon}
\end{figure}

\begin{figure*}
\includegraphics[width=0.4\columnwidth]{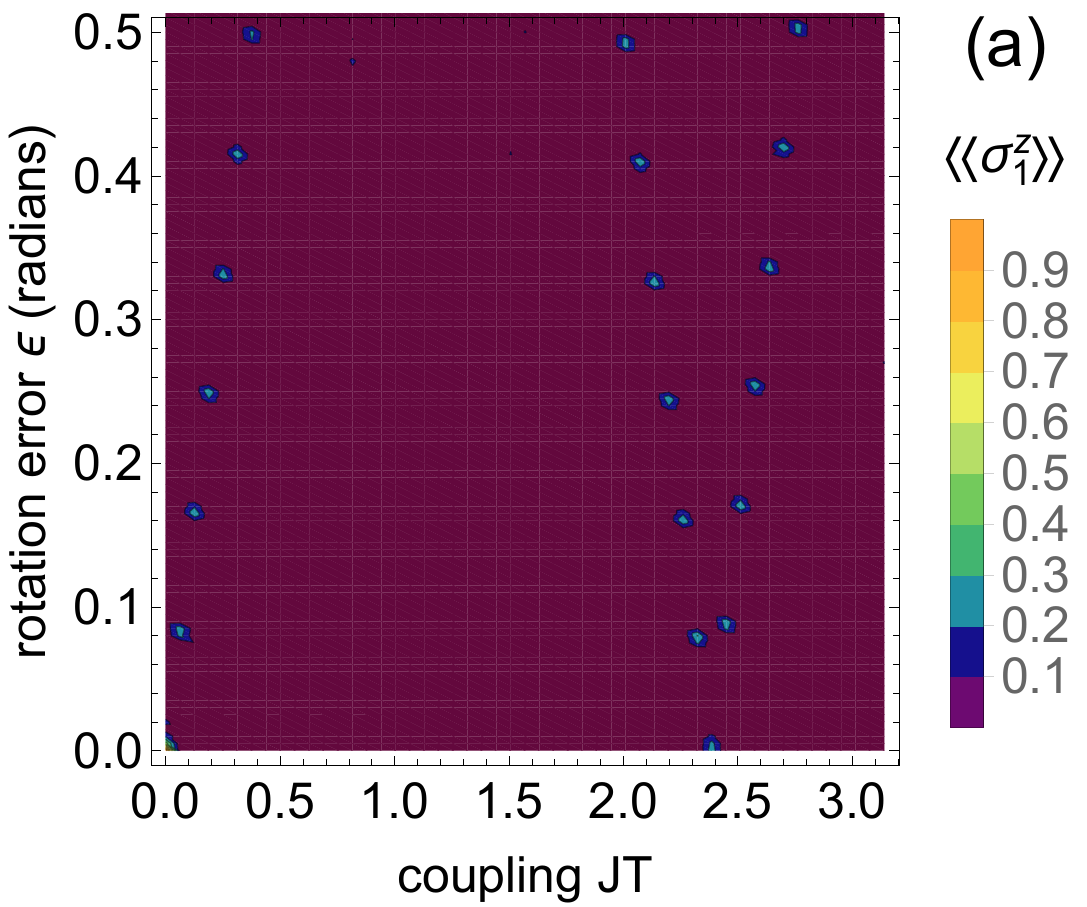}
\includegraphics[width=0.4\columnwidth]{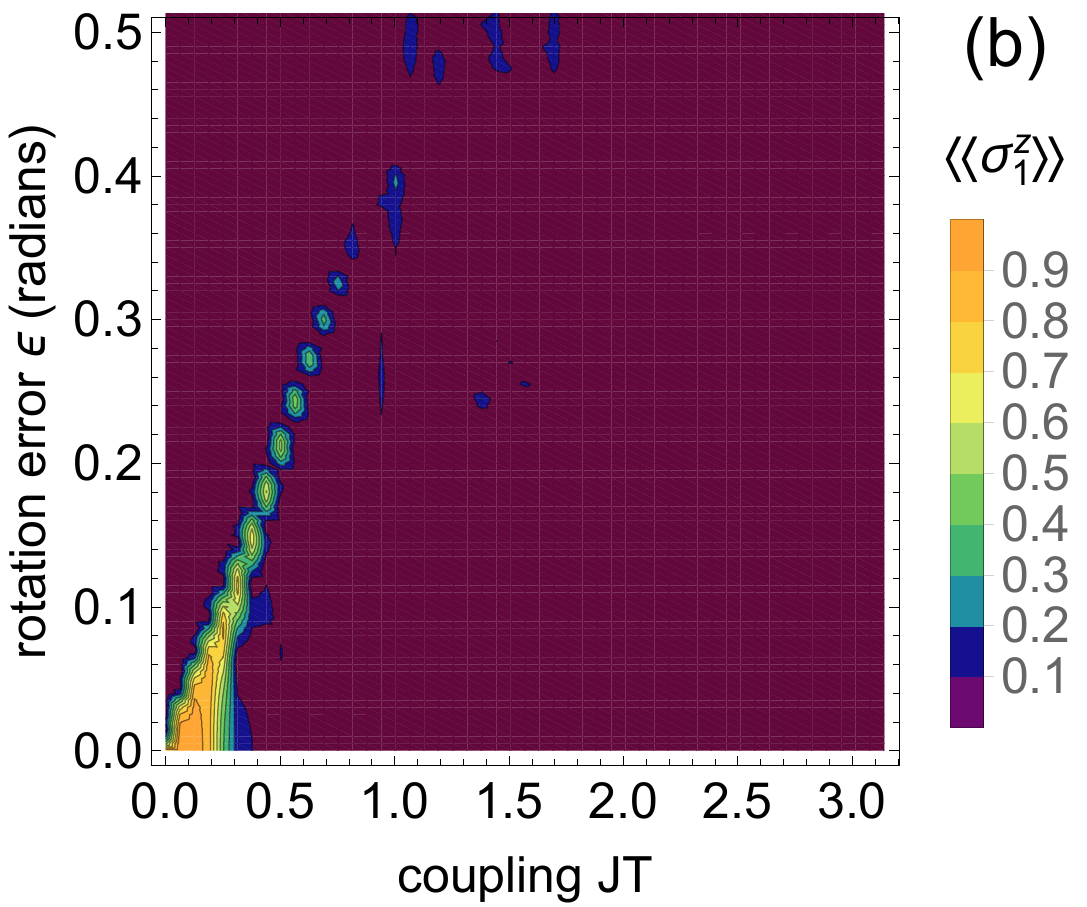}
\includegraphics[width=0.4\columnwidth]{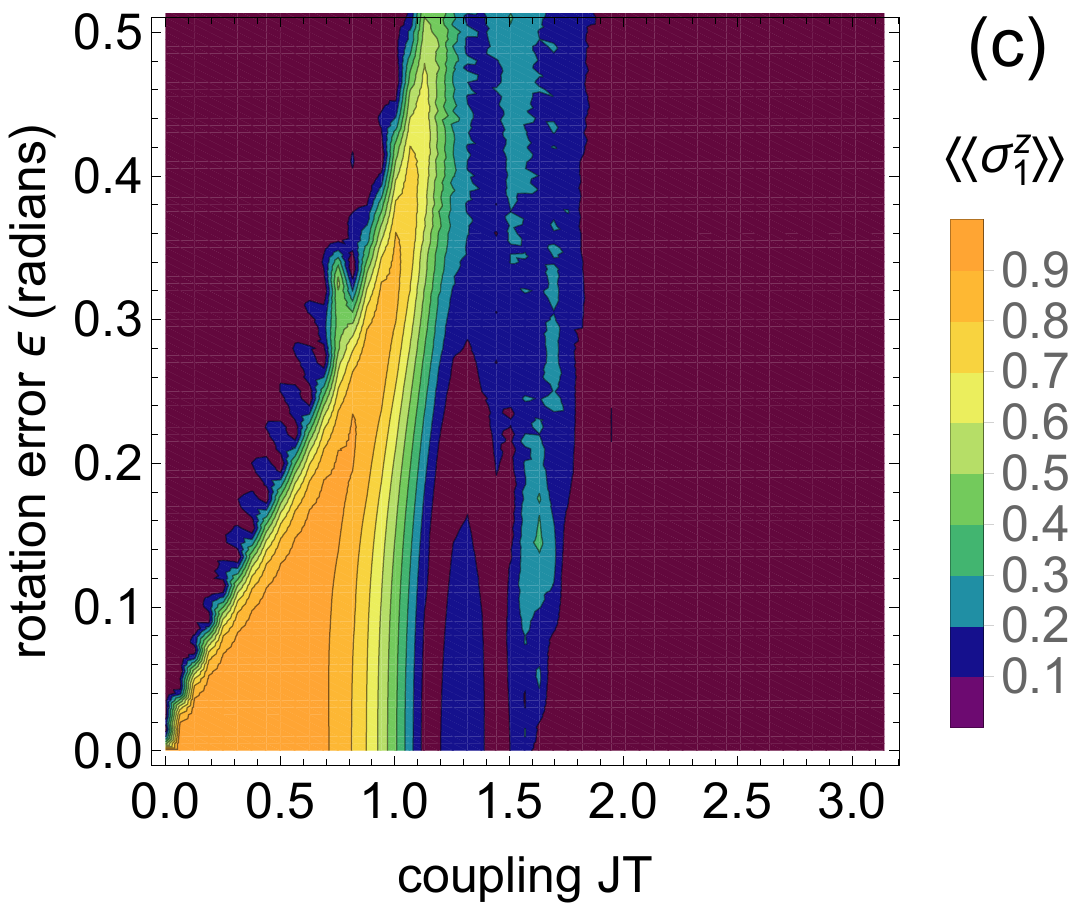}
\includegraphics[width=0.4\columnwidth]{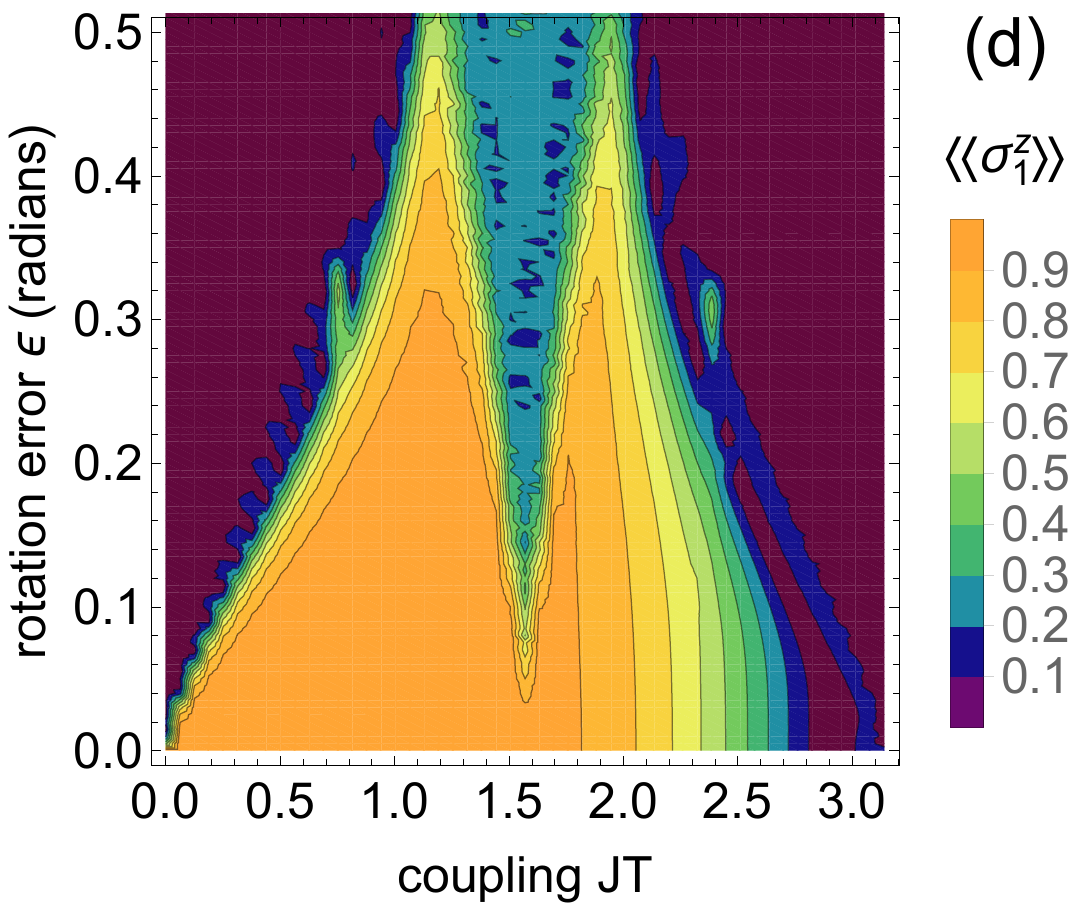}
\includegraphics[width=0.4\columnwidth]{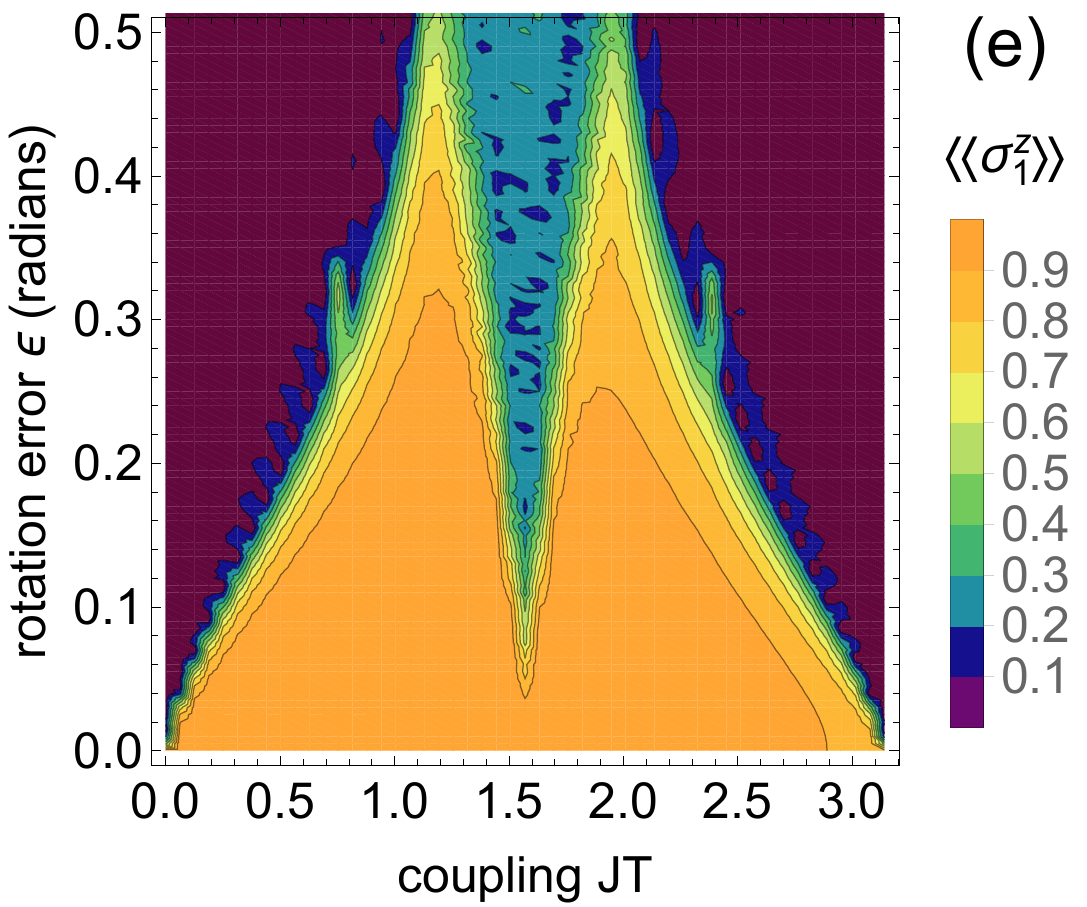}
\caption{Effect of H2I pulses. Phase diagrams for $N=4$ Heisenberg spin chain driven by one Floquet pulse about $x$ and (a) 0, (b) 2, (c) 16, (d) 64, (e) 128 H2I pulses about $z$ per period. The initial state is $\ket{\uparrow\downarrow\uparrow\downarrow}$, and the parameters are $h^zT=0.05$, $\delta h^zT=0.05$, $\delta JT=0$.}\label{fig:addingH2Ipulses}
\end{figure*}

\begin{figure*}
\includegraphics[width=0.4\columnwidth]{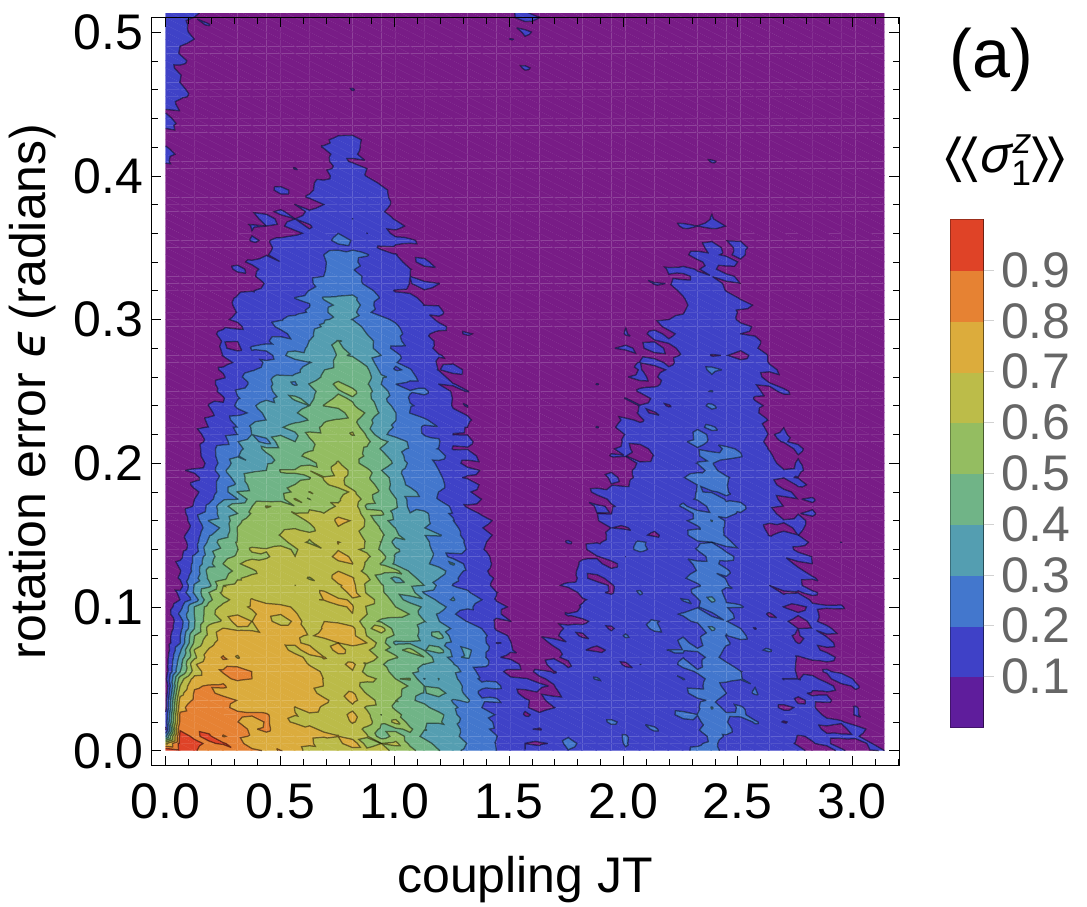}
\includegraphics[width=0.4\columnwidth]{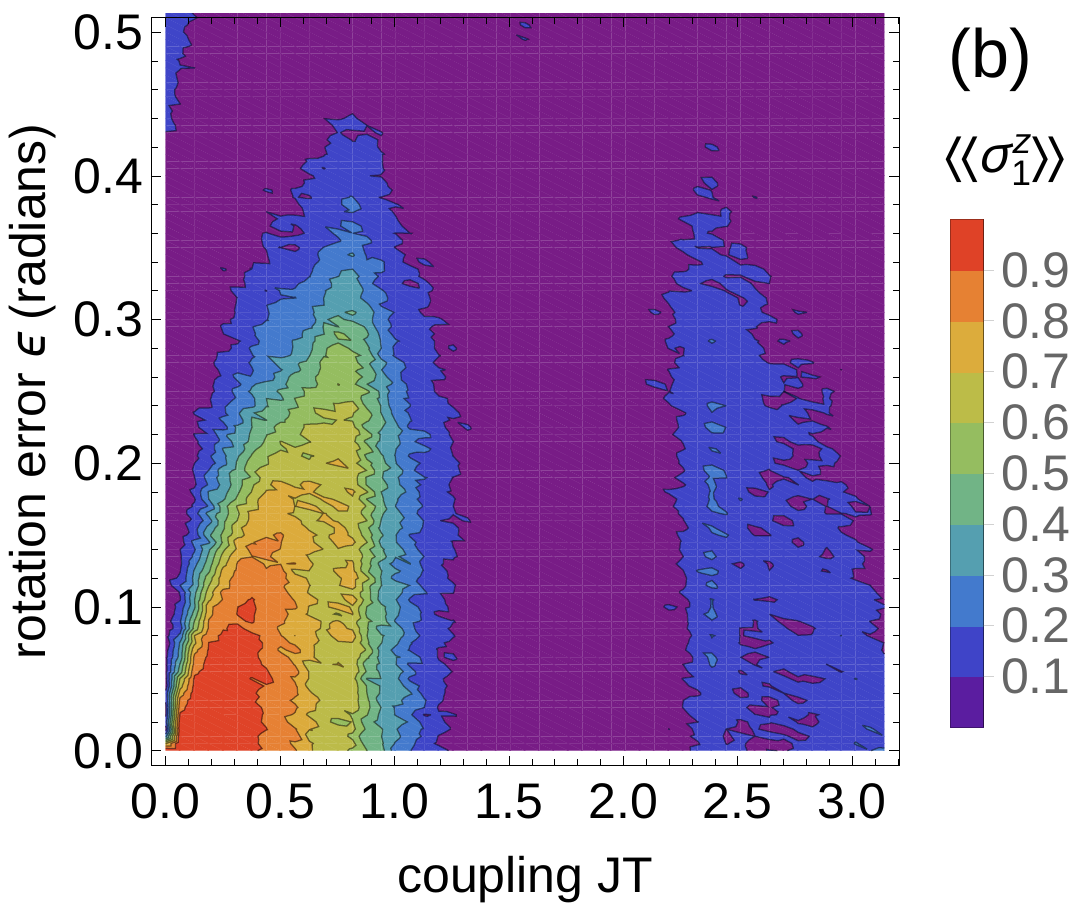}
\includegraphics[width=0.4\columnwidth]{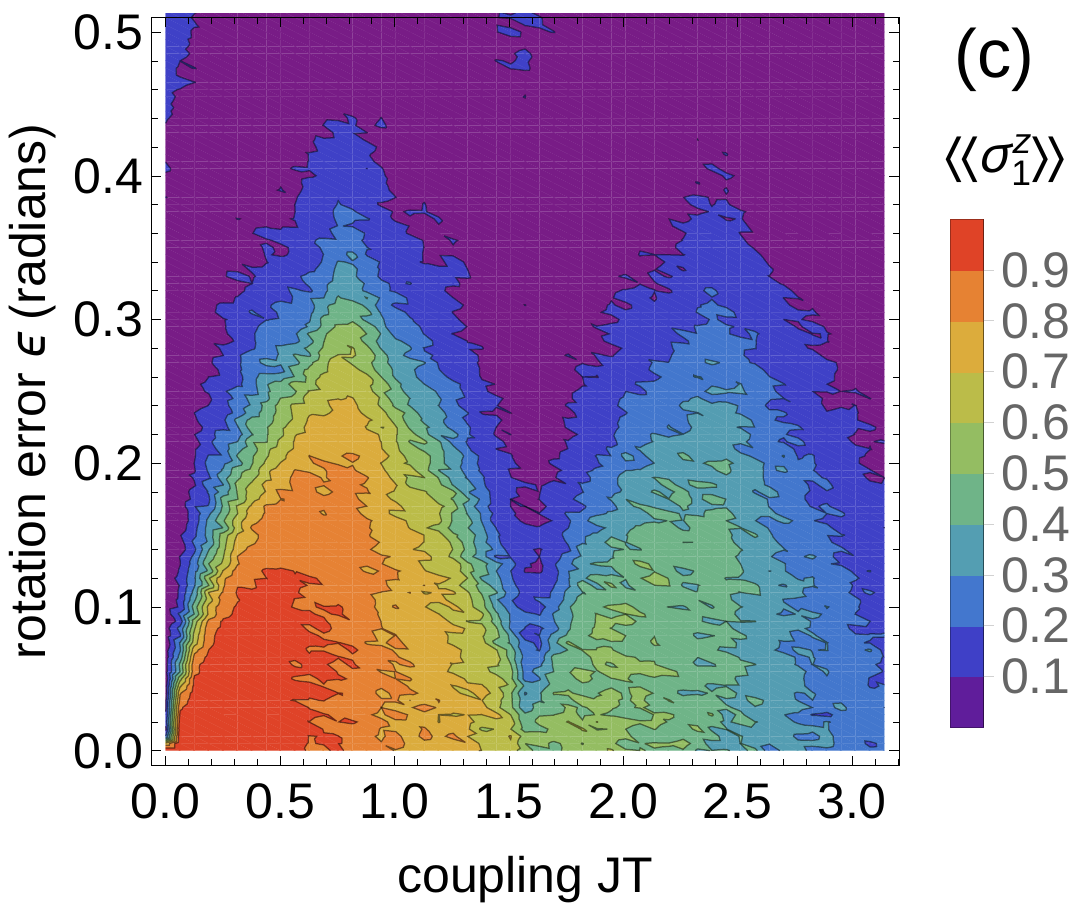}
\includegraphics[width=0.4\columnwidth]{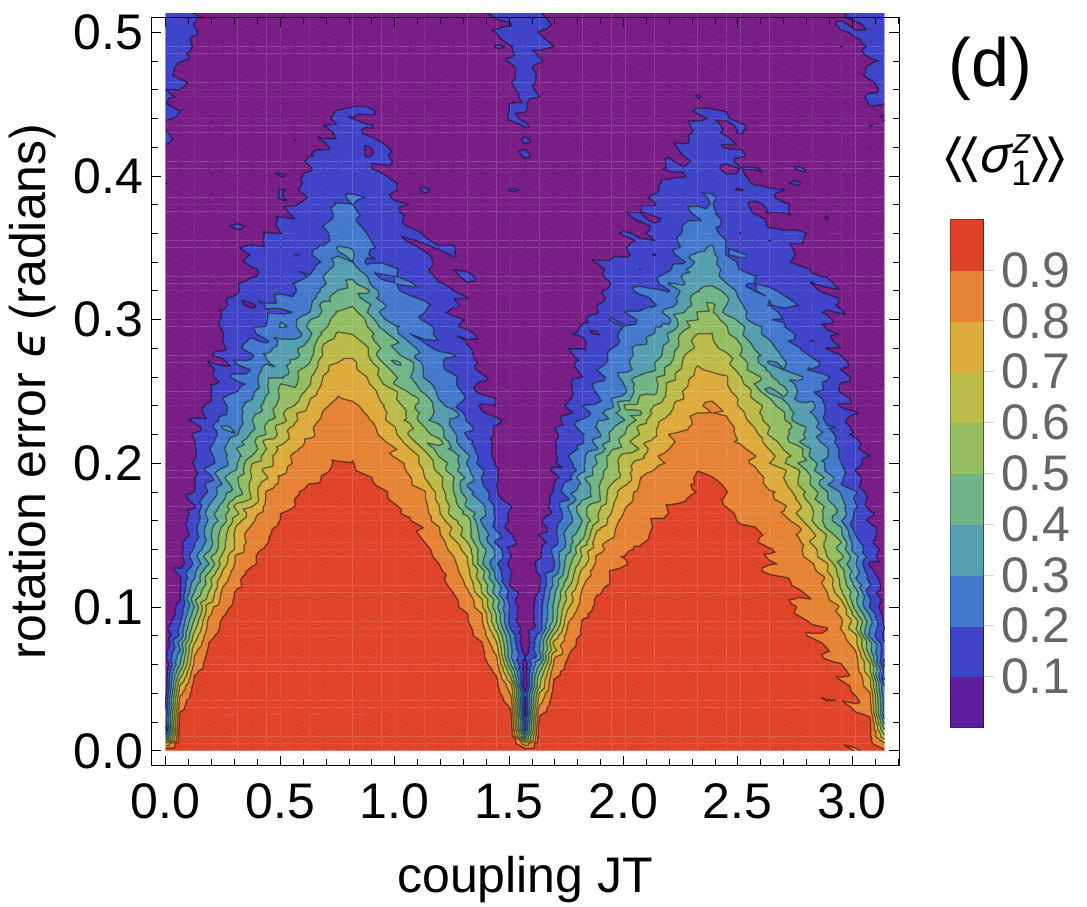}
\includegraphics[width=0.4\columnwidth]{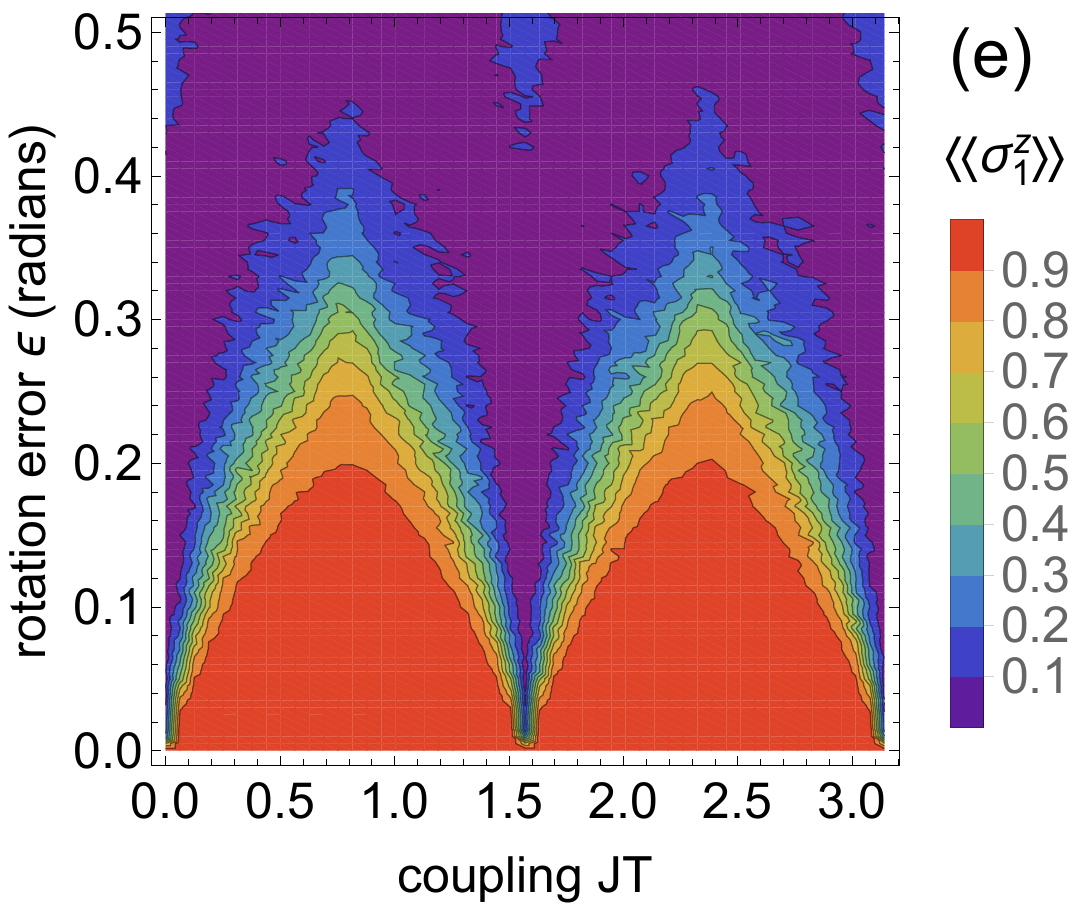}
\caption{Effect of H2I pulses for experimental parameter values. Phase diagrams for $N=4$ Heisenberg spin chain driven by one Floquet pulse about $x$ and (a) 0, (b) 2, (c) 16, (d) 64, (e) 128 H2I pulses about $z$ per period. The initial state is $\ket{\uparrow\downarrow\uparrow\downarrow}$, and the parameters are $h^zT=2\times10^4$, $\delta h^zT=50$, $\delta JT=0$.}\label{fig:addingH2Ipulses2}
\end{figure*}

We first consider the case where only one $\pi$ pulse about $x$ is applied during each drive period as in the Ising case above, so that the total Hamiltonian is $H=H_{Heisenberg}+H_{drive}(t)$, where
\begin{equation}
H_{drive}(t)=-(\pi/2-\epsilon)\sum_{k=1}^\infty\delta(t-kT)\sum_{i=1}^N\sigma_i^x.\label{pipulses}
\end{equation}
In the absence of coupling or local transverse field disorder, there is no discernible time crystal phase in this case. This is shown in Fig.~\ref{fig:addingH2Ipulses}(a), where we use the same external field and noise parameters as in Fig.~\ref{fig:isingpd1}(a) to make a direct comparison to the pure Ising case considered previously; we will return to the more physical parameter regime relevant for quantum dots shortly. Thus, we conclude that including spin-spin interactions along the $x$ or $y$ directions destroys the time crystal phase.

We now show that a time crystal phase can be recovered in the Heisenberg case by including additional pulses during each period. The basic idea is to exploit the following identity:
\begin{equation}
(\sigma_1^\beta\otimes{\mathbf 1})e^{i\theta\sum_\alpha\sigma_1^\alpha\otimes\sigma_2^\alpha}(\sigma_1^\beta\otimes{\mathbf 1})e^{i\theta\sum_\alpha\sigma_1^\alpha\otimes\sigma_2^\alpha}=e^{i2\theta\sigma_1^\beta\otimes\sigma_2^\beta},\label{H2Iidentity}
\end{equation}
which shows that a Heisenberg interaction can effectively be turned into an Ising interaction by applying $\pi$ pulses to every other spin at appropriate times. We will refer to such pulses as H2I pulses. The direction of the Ising interaction is determined by the rotation axis of the H2I pulses. Including two H2I pulses per period (in addition to the Floquet pulse) and setting $\theta=T/2$ exactly reproduces the Floquet operator of the Ising model, Eq.~\eqref{isingham}, in the absence of Floquet pulse errors. When these errors are present, the Floquet operator deviates from the pure Ising model case; however, the H2I pulses still produce a robust time crystal phase as shown in Fig.~\ref{fig:addingH2Ipulses}(b), although the range of spin-spin couplings and rotation errors over which this phase persists is considerably reduced compared to the pure Ising case shown in Fig.~\ref{fig:isingpd1}. However, the inclusion of additional H2I pulses increases the time crystal region, as is illustrated in Figs.~\ref{fig:addingH2Ipulses}(c)-(e). Here, the Floquet operator for $n$ H2I pulses has the form,
\begin{eqnarray}
{\cal U}_{F}&=&\left[e^{i\pi/2(\sigma_1^z+\sigma_3^z)}{\cal U}_H(T/n)e^{-i\pi/2(\sigma_1^z+\sigma_3^z)}{\cal U}_H(T/n)\right]^{n/2}\nn\\&\times&e^{i(\pi/2-\epsilon)\sum_i\sigma_i^x},
\end{eqnarray}
where ${\cal U}_H(T/n)=e^{-iH_{Heisenberg}T/n}$, and we are considering an $N=4$ spin chain. The full sequence of pulses applied in a single period is illustrated in Fig.~\ref{fig:H2Icartoon}. Note that we only apply H2I pulses on spins 1 and 3 in conformity with the identity given in Eq.~\eqref{H2Iidentity}. We see from Fig.~\ref{fig:addingH2Ipulses}(e) that for a sufficiently large number of H2I pulses per period, in this case $n=128$, the size of the time crystal phase region saturates, and the phase diagram for the pure Ising case is recovered.

The H2I pulses have a similar effect in the parameter regime relevant for GaAs quantum dot arrays, as illustrated in Fig.~\ref{fig:addingH2Ipulses2}. Here, we take $h^zT=2\times10^4$ and $\delta h^zT=50$, which correspond to an external magnetic field of approximately 0.5 T and an Overhauser noise level of about 8 MHz for a pulse period of $T=1 \mu$s. Henceforth, we use a different color scale for phase diagrams to distinguish the GaAs parameter regime from the other regimes considered in Figs.~\ref{fig:isingpd1} and \ref{fig:addingH2Ipulses}. We again see that a large time crystal phase region emerges as the number of H2I pulses per period increases. In Appendix~\ref{app:DTTSB}, we confirm that the discrete time translation symmetry is indeed broken by showing that the time evolution of the spin vector in these regions is not $T$-periodic. We have also investigated the dependence of this phase region on the strength of the external magnetic field but found essentially no change as the field is reduced to $h^z=0$. This is consistent with similar findings in earlier studies of the MBL phase in short Heisenberg spin chains, where the insensitivity to the external magnetic field can be attributed to the conservation of total $S^z$ \cite{Barnes_PRB16}. As shown in Appendix~\ref{app:isingNdependence}, for experimentally relevant levels of charge noise, the effect on the time crystal phase is negligible, so we henceforth set $\delta J=0$. The diagrams in Fig. 7 also remain essentially unchanged when we include the same amount of pulse error in the H2I pulses as in the Floquet pulses.

\begin{figure}
\includegraphics[width=0.49\columnwidth]{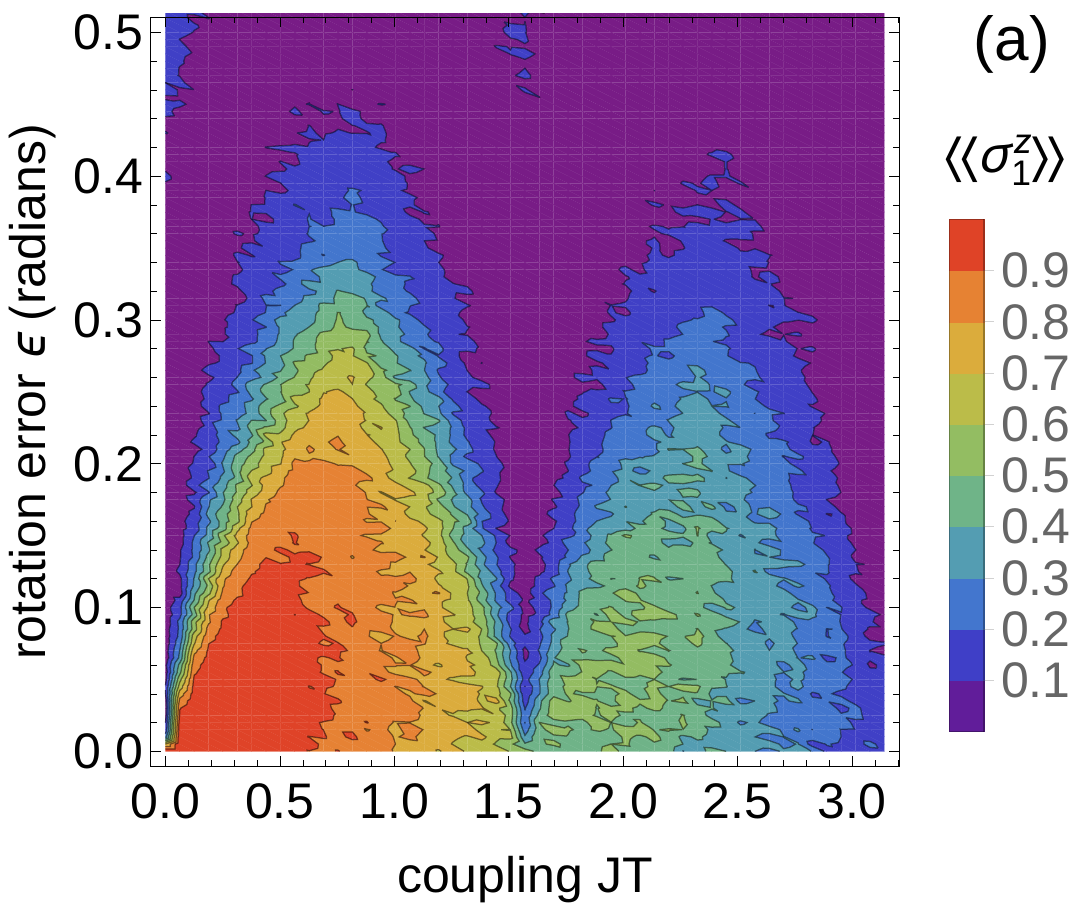}
\includegraphics[width=0.49\columnwidth]{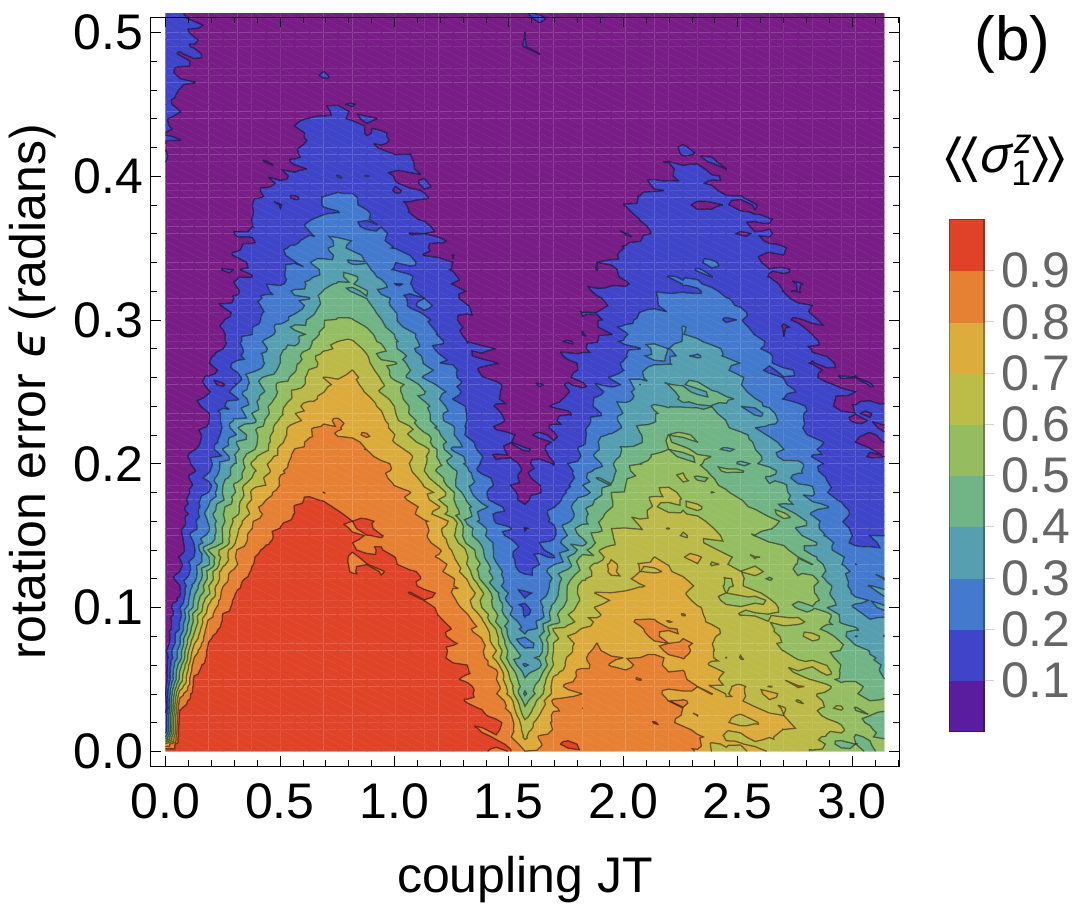}
\includegraphics[width=0.49\columnwidth]{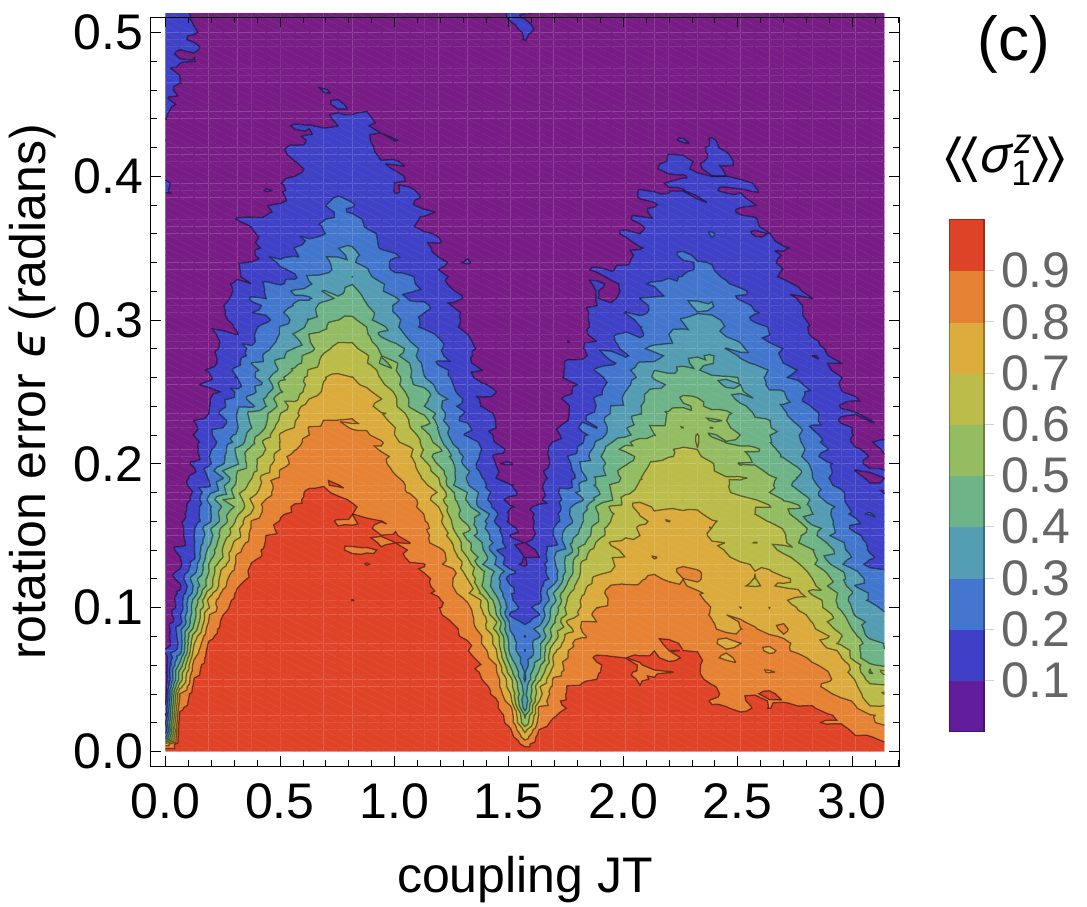}
\includegraphics[width=0.49\columnwidth]{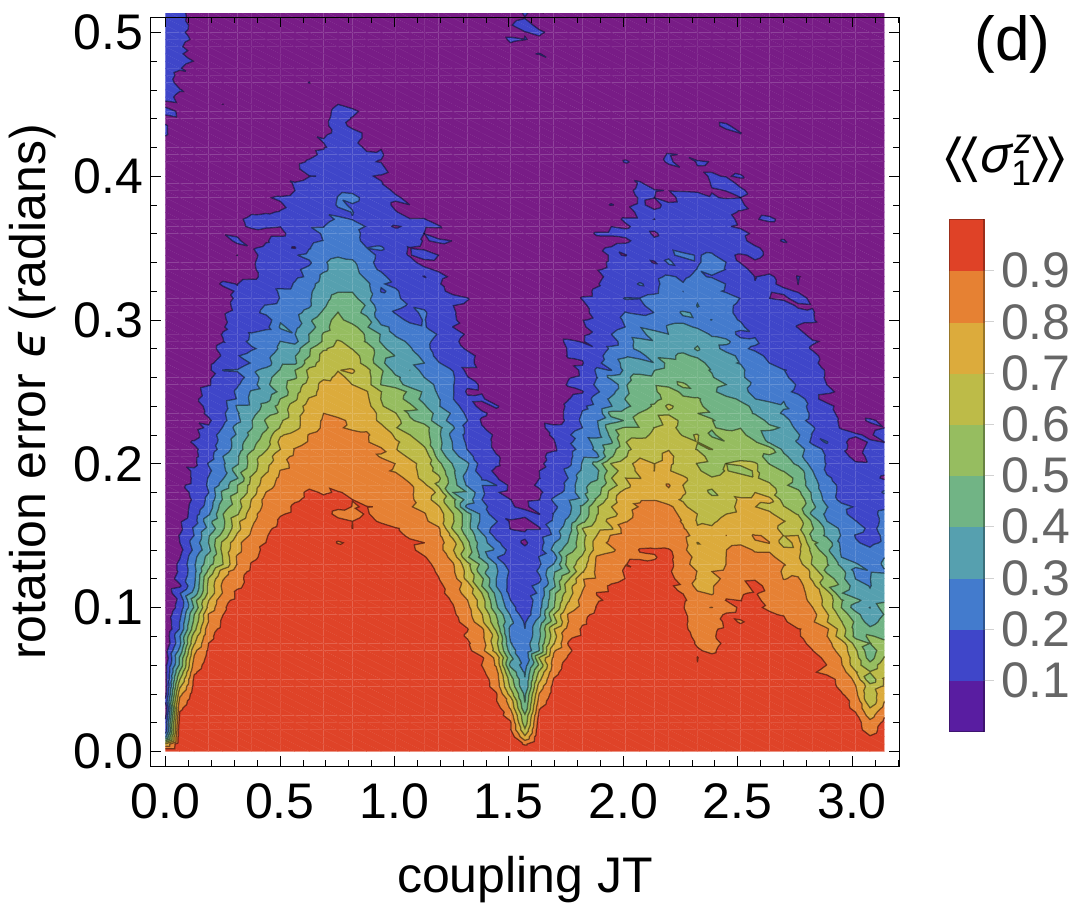}
\caption{Dependence on initial state. Phase diagrams for Heisenberg spin chains driven by one Floquet pulse about $x$ and 16 H2I pulses about $z$ per period. The initial states are (a) $\ket{\uparrow\downarrow\downarrow\uparrow}$, (b) $\ket{\uparrow\uparrow\downarrow\downarrow}$, (c) $\ket{\uparrow\uparrow\uparrow\downarrow}$, (d) $\ket{\uparrow\uparrow\uparrow\uparrow}$, and the system parameters are $h^zT=2\times10^4$, $\delta h^zT=50$, $\delta JT=0$.}\label{fig:heisenberg_changinginitstate}
\end{figure}

Phase diagrams for different initial states for a Heisenberg spin chain is shown in Fig.~\ref{fig:heisenberg_changinginitstate}. We see that, unlike the pure Ising case, the time-averaged spin component, $\mean{\mean{\sigma_1^z}}$, is sensitive to the initial state. Among the $N=4$ states with total spin $S^z=0$, the N\'eel-ordered state exhibits the least robust time crystal behavior (c.f. Fig.~\ref{fig:addingH2Ipulses2}(c)), while the initial state $\ket{\uparrow\uparrow\downarrow\downarrow}$ is the most robust. This can be understood from the fact that the exchange interaction only affects adjacent pairs of spins that have opposite orientations. Because the state $\ket{\uparrow\uparrow\downarrow\downarrow}$ contains only one such pair of spins, the system remains close to this initial state for a broader range of couplings. When total $S^z\ne0$, we see that the time crystal phase becomes even more robust because of the reduced Hilbert space. 

\section{Spin rotations and coherence in the time crystal phase}\label{sec:control}

So far, we have seen that the time crystal phase preserves multi-spin states which are tensor products of single-spin states oriented either parallel or antiparallel to the (effective) Ising interaction and orthogonal to the Floquet pulses. In this section, we examine whether additional multi-spin states can also be preserved and whether interactions are on a whole beneficial for preserving arbitrary single-spin states. Regarding the latter, we can imagine using a spin at the end of the chain as a qubit and ask whether interactions with the other spins prolong the coherence time of this spin via time crystallization.

\begin{figure}
\includegraphics[width=0.49\columnwidth]{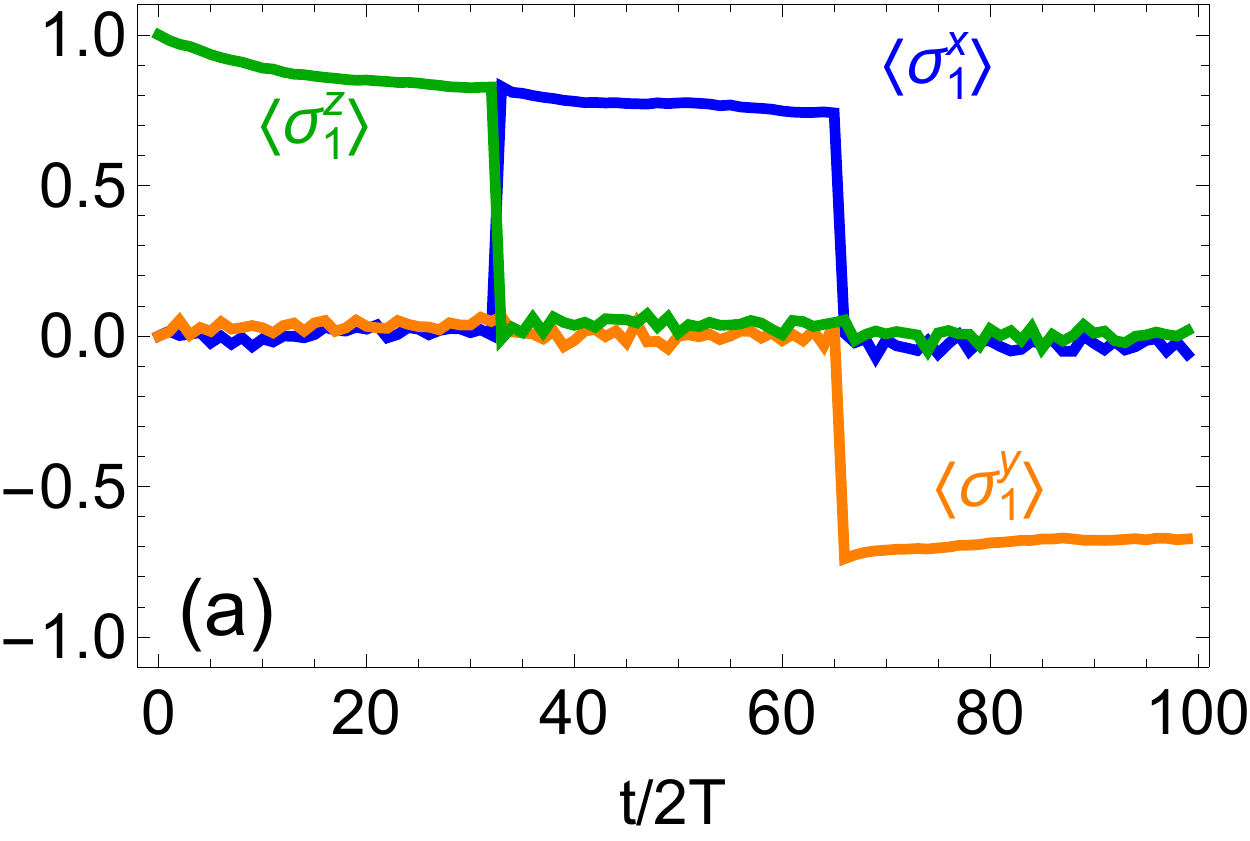}
\includegraphics[width=0.49\columnwidth]{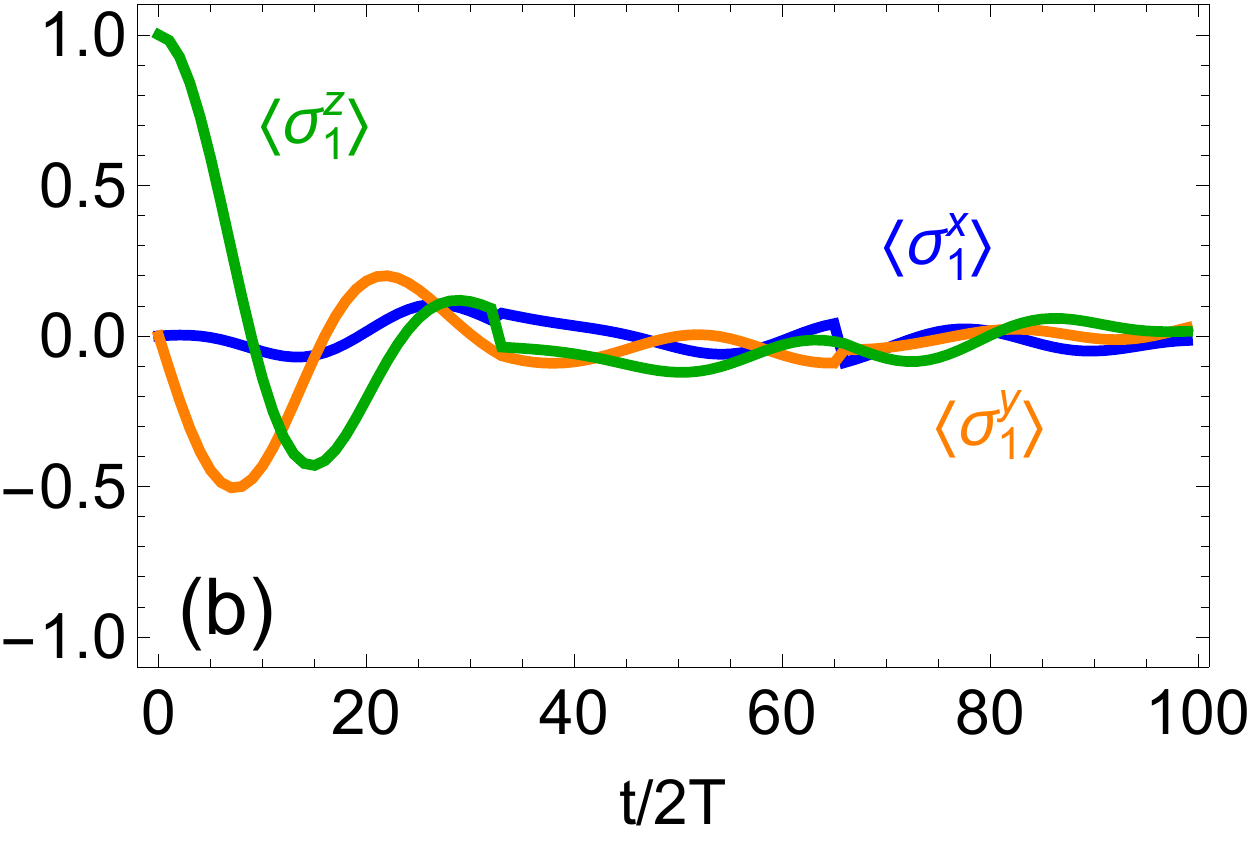}
\includegraphics[width=\columnwidth]{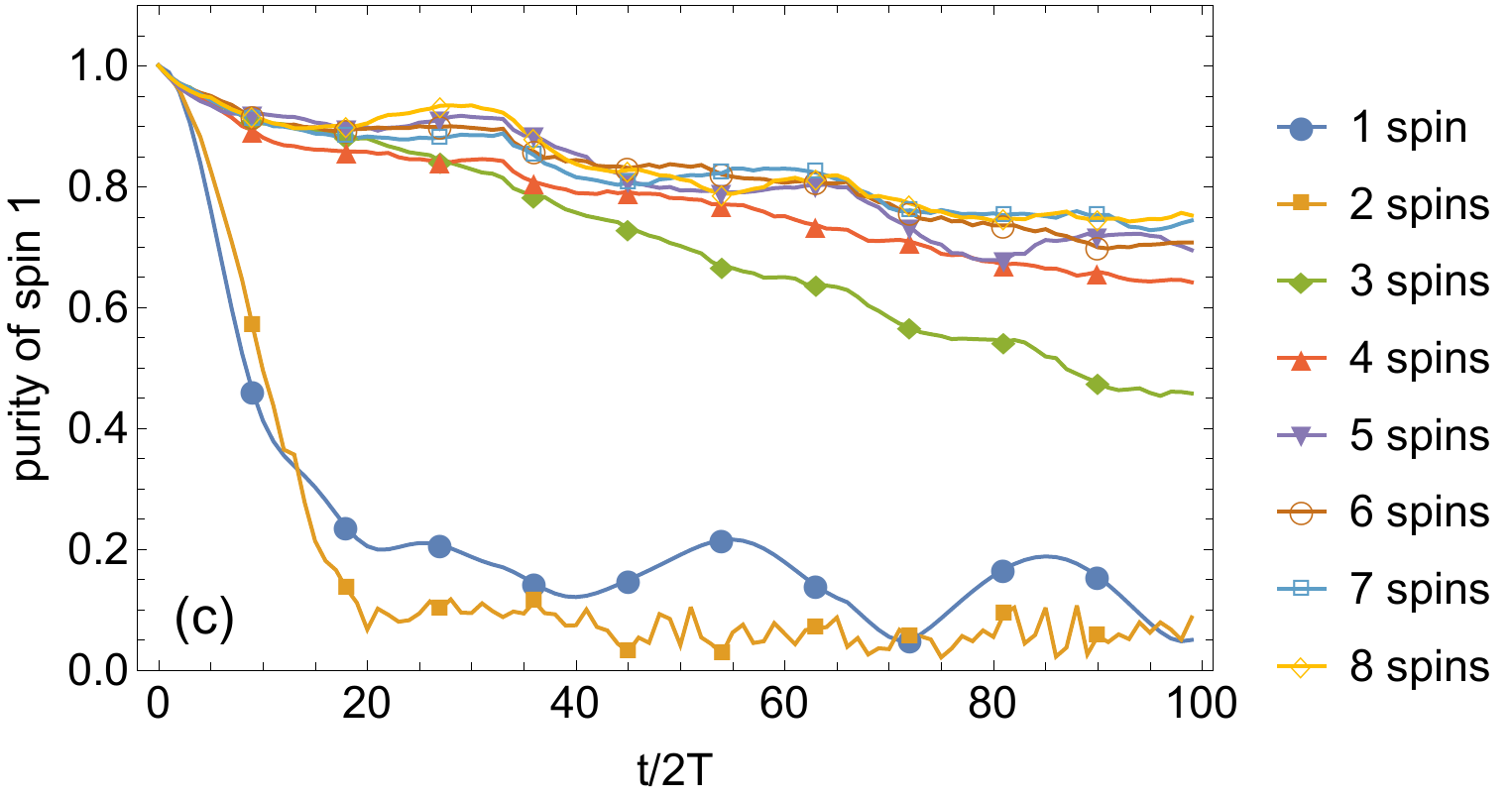}
\caption{Spin rotations in the time crystal phase. (a), (b) Spin vector components for a spin at the end of a $N=4$ Heisenberg spin chain prepared in the state $\ket{\uparrow\uparrow\uparrow\uparrow}$. One Floquet pulse along $x$ and 128 H2I pulses along $z$ are applied per period. At $t=66T$, a $\pi/2$ rotation about $y$ is applied to all spins, bringing the spins into the state $\ket{\uparrow_x\uparrow_x\uparrow_x\uparrow_x}$. Afterward, the Floquet and H2I pulse axes are rotated to $y$ and $x$, respectively. At $t=132T$, a $\pi/2$ rotation about $z$ is applied, bringing the spins into the state $\ket{\downarrow_y\downarrow_y\downarrow_y\downarrow_y}$. Afterward, the Floquet and H2I pulse axes are rotated to $z$ and $y$, respectively. (c) Purity of the end spin under the same rotations for different numbers $N$ of spins in the chain prepared in a tensor product of $\ket{\uparrow}$. For all panels, the system and noise parameters are $h^zT=0$, $\delta h^xT=\delta h^yT=\delta h^zT=10$, $\epsilon=0.05$, $\delta JT=0$, and (a) $JT=\pi$, (b) $J=0$. The H2I pulses also include a rotation angle error of size $\epsilon$.}\label{fig:spinrotations}
\end{figure}

\begin{figure}
\includegraphics[width=0.49\columnwidth]{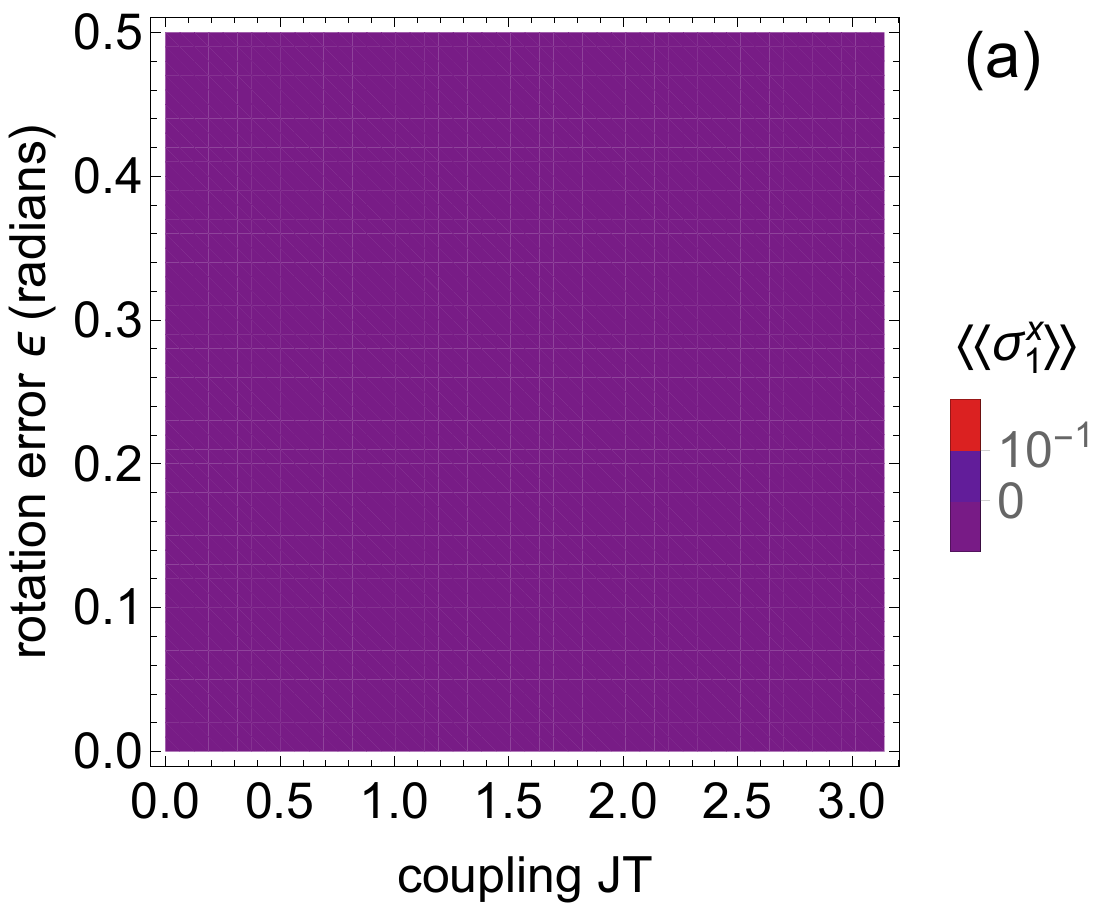}
\includegraphics[width=0.49\columnwidth]{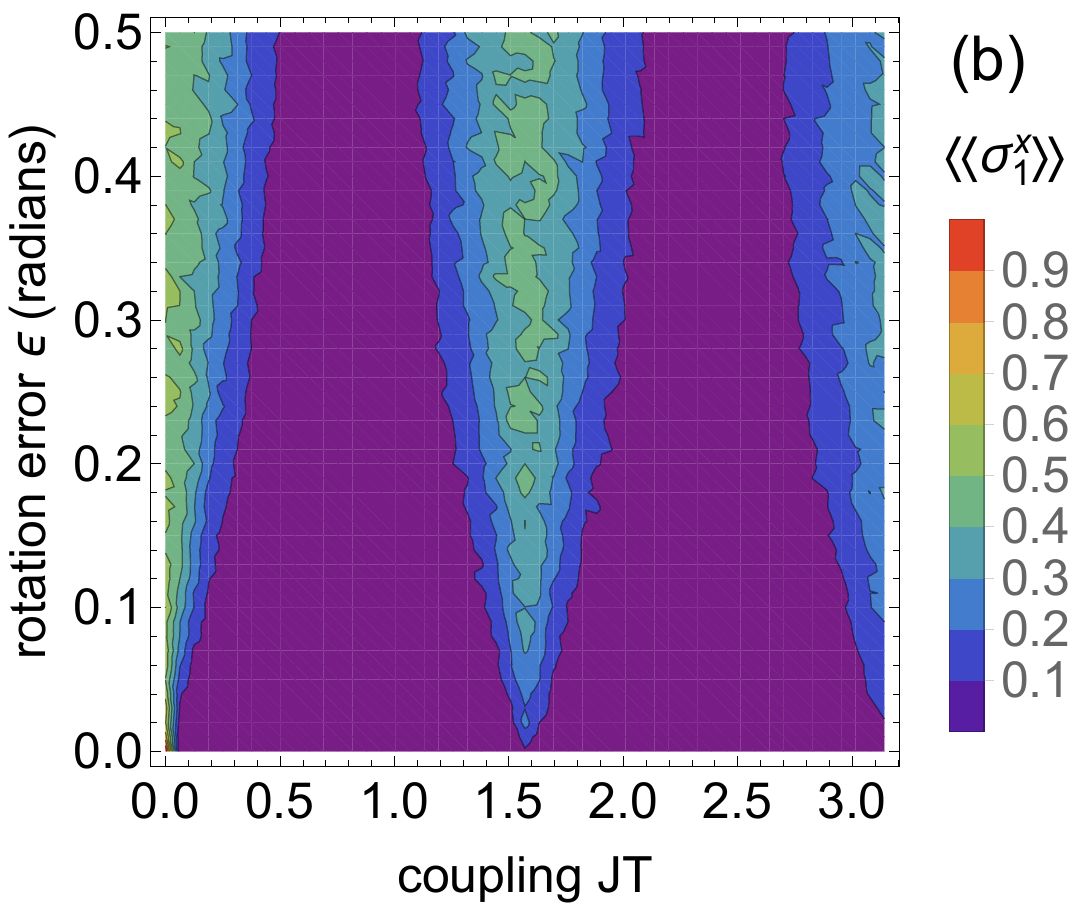}
\includegraphics[width=0.49\columnwidth]{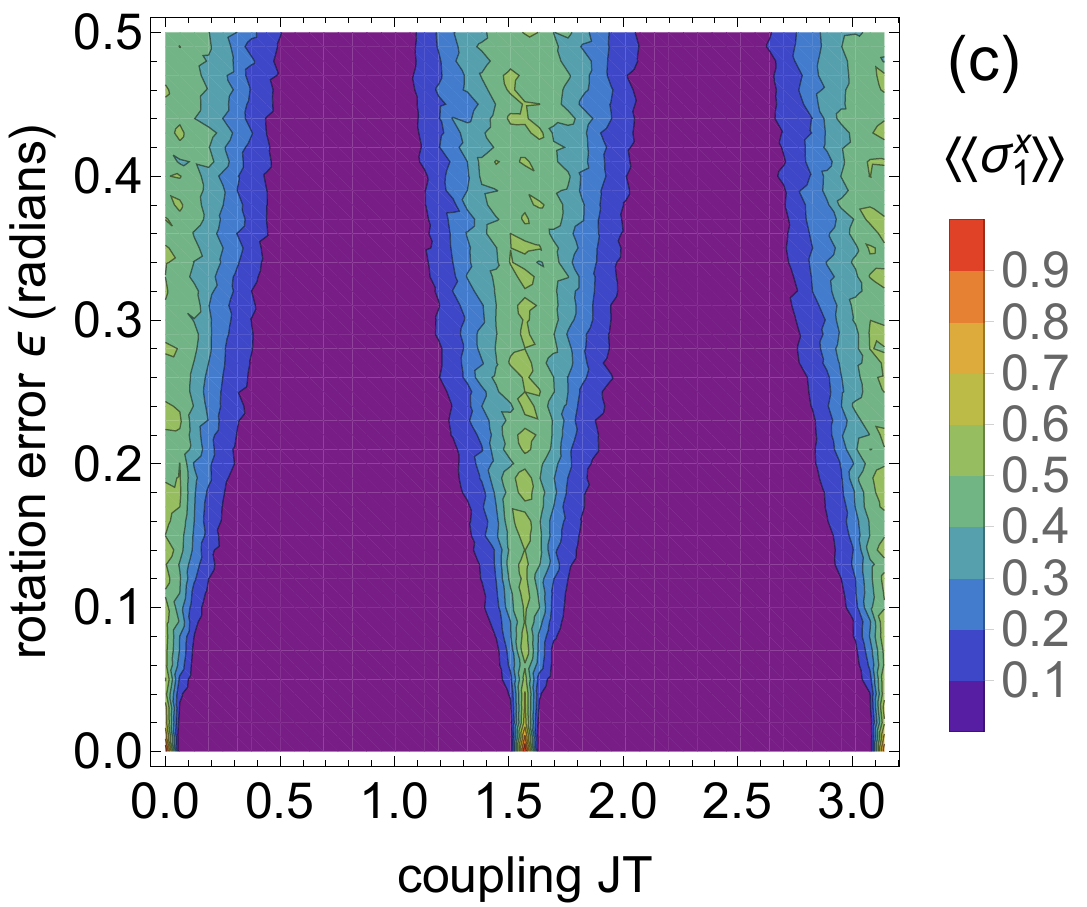}
\includegraphics[width=0.49\columnwidth]{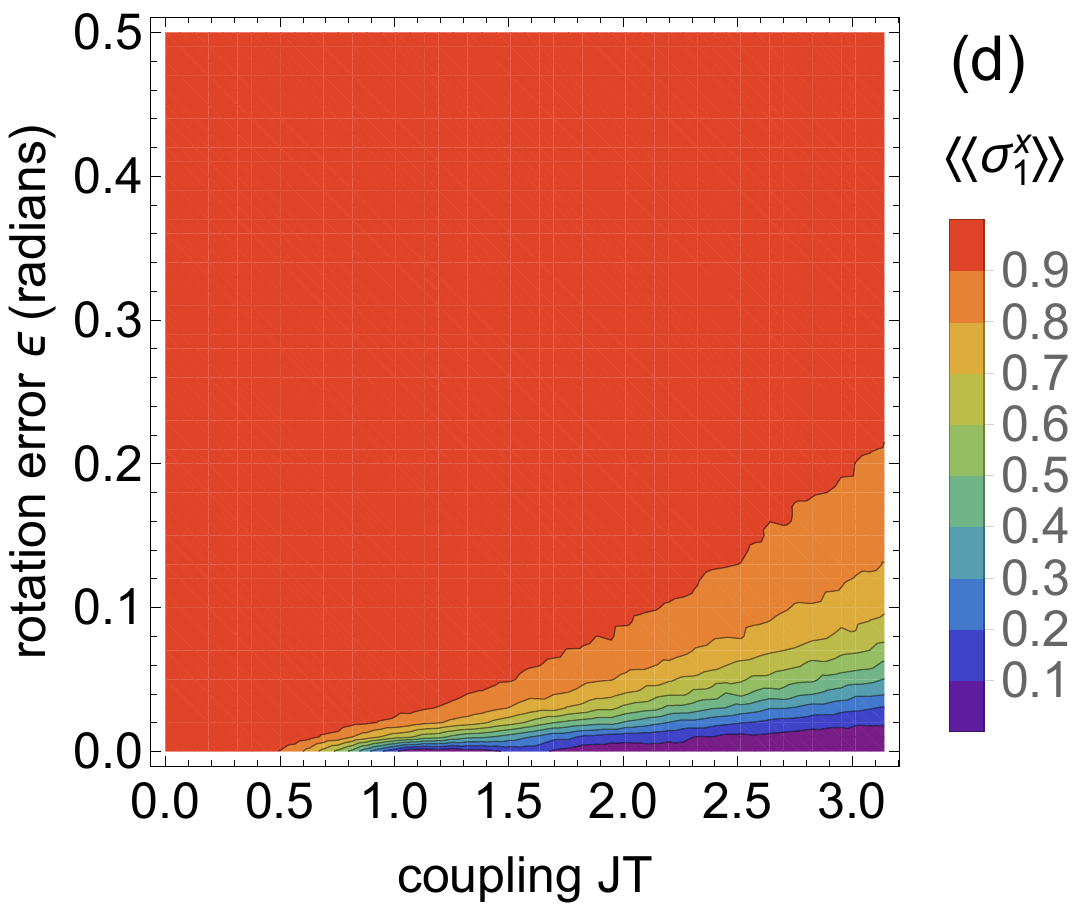}
\caption{Preservation of spin coherence. $\mean{\mean{\sigma_1^x}}$ phase diagrams for Heisenberg spin chains initialized in state $\ket{\uparrow_x\uparrow_x\uparrow_x\uparrow_x}$ and driven by (a) no pulses, (b) one Floquet pulse about $x$, (c) one Floquet pulse about $x$ and 128 H2I pulses about $z$, (d) one Floquet pulse about $x$ and 128 H2I pulses about $x$ per period. The system parameters are $h^zT=2\times10^4$, $\delta h^zT=50$, $\delta JT=0$.}\label{fig:heisenberg_coherence}
\end{figure}

The fact that we can select the orientation of the effective Ising interaction by changing the rotation axis of the H2I pulses suggests that we can dynamically switch between different time crystal phases and thus preserve different sets of multi-spin states. This is indeed the case, as is demonstrated in Fig.~\ref{fig:spinrotations}(a). Here, we prepare the spin chain in the state $\ket{\uparrow\uparrow\uparrow\uparrow}$ (along the $z$ axis) and then apply two $\pi/2$ rotations at later times---the first about the $y$ axis, and the second about $z$. By adjusting the axes of the H2I pulses and Floquet pulses appropriately after each rotation, the state of the end spin remains mostly pure over many Floquet periods. This is true despite the inclusion of errors not only in the Floquet pulses, but also in the H2I pulses. Fig.~\ref{fig:spinrotations}(b) shows that interactions play a critical role in maintaining high purity (i.e., spin vector length $\sqrt{\mean{\sigma^x}^2+\mean{\sigma^y}^2+\mean{\sigma^z}^2}$) throughout these rotations and beyond: when $J=0$, all spin vector components quickly decay to zero. In addition, we have set the external magnetic field to zero, $h^z=0$, in Fig.~\ref{fig:spinrotations} because we find that if the field is not aligned with the H2I pulses, then it is detrimental to the time crystal phase. Of course, rotating the external field along with the pulse axes after each spin rotation would avoid this issue; however, as was mentioned above, the external field does not appear to provide any additional stability to the time crystal phase, so we set it to zero. In this case, there is no natural quantization axis for the spin chain, so we have included local field noise, $\delta h$, along all three directions. With this type of noise, we find that the optimal interaction strengths are no longer in the vicinity of $JT=\pi/2$ and $3\pi/2$ (as in Fig.~\ref{fig:addingH2Ipulses2}), but instead around $JT=\pi$. In Fig.~\ref{fig:spinrotations}(c), we consider the spin vector evolution under the same two $\pi/2$ rotations as in panels (a) and (b), but now with varying numbers of spins in the chain. The figure shows the purity of the end spin for spin chain lengths of $N=1,...,8$, where again the chain is prepared in a tensor product state with each spin in $\ket{\uparrow}$. The purity rapidly increases as $N$ is increased from 1 to 4. Some additional improvement results from increasing $N$ further up to 5 or 6 spins, but then the purity essentially saturates beyond this, with only very modest gains resulting from an increase up to $N=8$ spins. Thus, it seems that the end spin is not sensitive to spins beyond 5 or 6 sites away.

Next, we return to the case where $h^z\ne0$ and the local field noise is only along the $z$ direction, and we now investigate the preservation of a product of superposition states: $\ket{\uparrow_x\uparrow_x\uparrow_x\uparrow_x}$, where $\ket{\uparrow_x}=(\ket{\uparrow}+\ket{\downarrow})/\sqrt{2}$, to see whether time crystals can preserve spin coherence. In the absence of any Floquet or H2I pulses, the state quickly dephases, as shown in Fig.~\ref{fig:heisenberg_coherence}(a). Including Floquet pulses along the $x$ direction slows down this dephasing (see Fig.~\ref{fig:heisenberg_coherence}(b)), which is of course expected since the Floquet pulses implement dynamical decoupling. Here, the decoupling is imperfect because of the pulse errors. Interestingly, this decoupling happens not only in the noninteracting case $J=0$, but also when the coupling strength is close to $JT=\pi/2$. Away from these special values of $J$, interactions interfere with the dynamical decoupling, resulting in fast dephasing. This interference happens regardless of whether the interactions are of Heisenberg or Ising type, so including H2I pulses along $z$ has little effect in this case (Fig.~\ref{fig:heisenberg_coherence}(c)). However, if we switch the H2I pulse axis to $x$, then the dephasing is strongly suppressed, as shown in Fig.~\ref{fig:heisenberg_coherence}(d). This suppression cannot be attributed to a time crystal phase however, as is evident from the fact that dephasing speeds up monotonically as the interaction strength $J$ is increased. Instead, we attribute this behavior to the fact that the H2I and Floquet pulses combine to form a CPMG sequence, and this sequence is well known to be robust against pulse errors \cite{Meiboom_Gill}. 

\begin{figure}
\includegraphics[width=0.7\columnwidth]{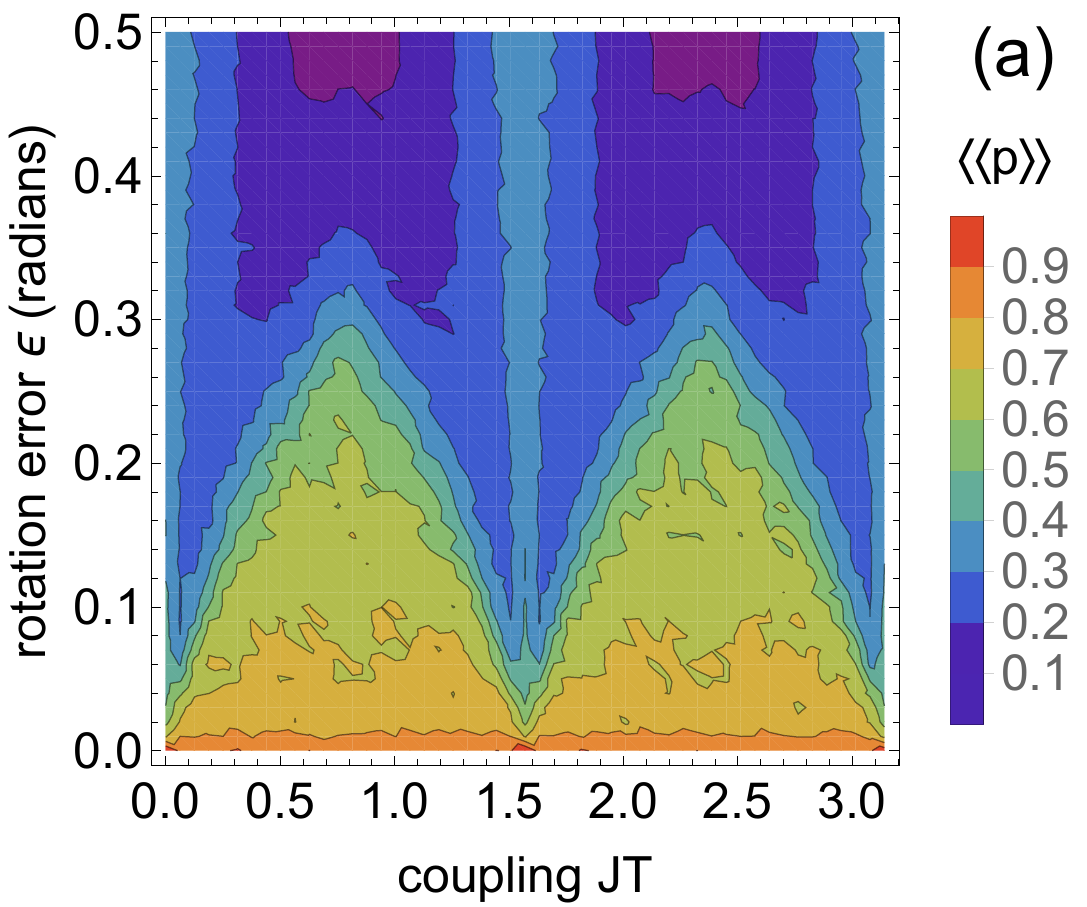}
\includegraphics[width=0.7\columnwidth]{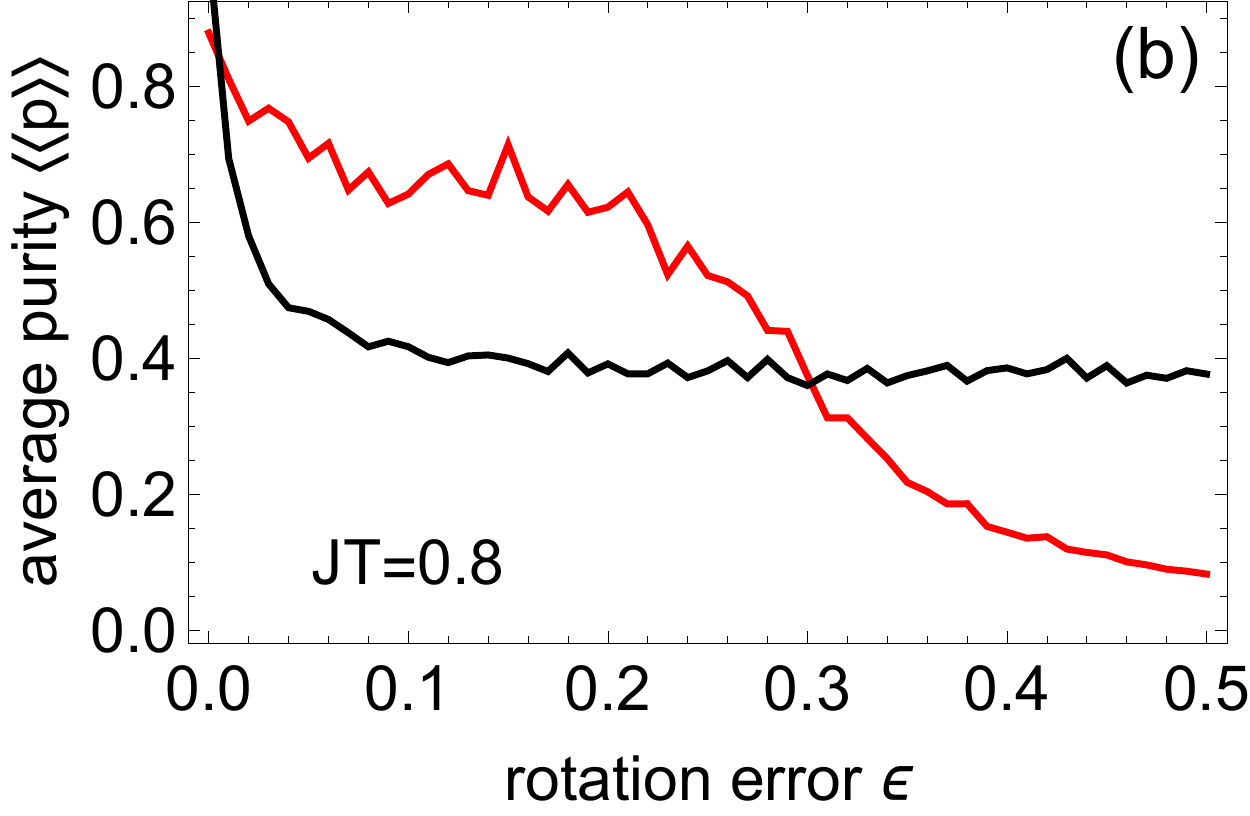}
\caption{Purity for $N=4$ Heisenberg spin chain averaged over 200 Floquet periods and over all initial states of the form $\ket{\psi\psi\psi\psi}$. One Floquet pulse about $x$ and 128 H2I pulses about $z$ are applied per period. The system parameters are $h^zT=2\times10^4$, $\delta h^zT=50$, $\delta JT=0$. (a) The full phase diagram. (b) A 1D cut of the phase diagram at $JT=0.8$ (red curve) and the corresponding result for $J=0$ (black curve).}\label{fig:purity}
\end{figure}

Thus, it would seem that interactions do not improve the performance of dynamical decoupling in preserving states like $\ket{\uparrow_x\uparrow_x\uparrow_x\uparrow_x}$ but in fact harm it. This does not mean, however, that time crystals do not provide any benefit in terms of state preservation. The findings illustrated in Fig.~\ref{fig:spinrotations} clearly show that {\it some} states can be preserved as a consequence of driving and interactions. The question then becomes whether most states benefit from the combined effect of these elements. To address this question, we compute the purity $\mean{\mean{p}}$ of the end spin averaged over all initial multi-spin states of the form $\ket{\psi\psi\psi\psi}$, where $\ket{\psi}$ is any single-spin state: $\ket{\psi}=\cos\theta\ket{\uparrow}+\sin\theta e^{i\chi}\ket{\downarrow}$ with arbitrary $\theta$ and $\chi$. We also average over noise and over 200 Floquet periods. The results are shown in Fig.~\ref{fig:purity}. Panel (a) shows a phase diagram for the average purity, where it is evident that turning on interactions increases the purity in the presence of pulse errors. The structure of this phase diagram is reminiscent of the phase diagrams for the spin vector component shown in Fig.~\ref{fig:addingH2Ipulses2}. In particular, the preservation is strongest near $JT=\pi/4$ and $3\pi/4$. Panel (b) shows a one-dimensional cut at $JT=0.8$ (red curve) in comparison with the result for no interactions (black curve). It is clear that for a range of pulse errors extending from $\epsilon\approx0.005$ up to $\epsilon\approx0.3$, the presence of interactions improves the average purity substantially, while for pulse errors outside of this window, interactions have a negative effect. This suggests that driving and interactions together may provide a net benefit towards quantum state preservation in the regime of realistic pulse errors and system parameters.

\begin{figure}
\includegraphics[width=0.75\columnwidth]{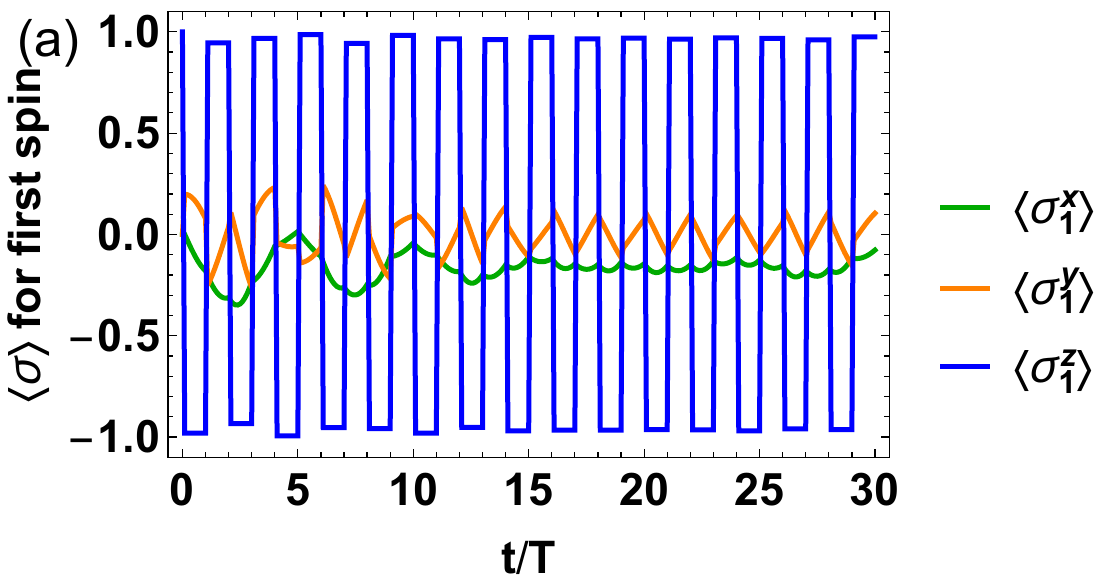}
\includegraphics[width=0.75\columnwidth]{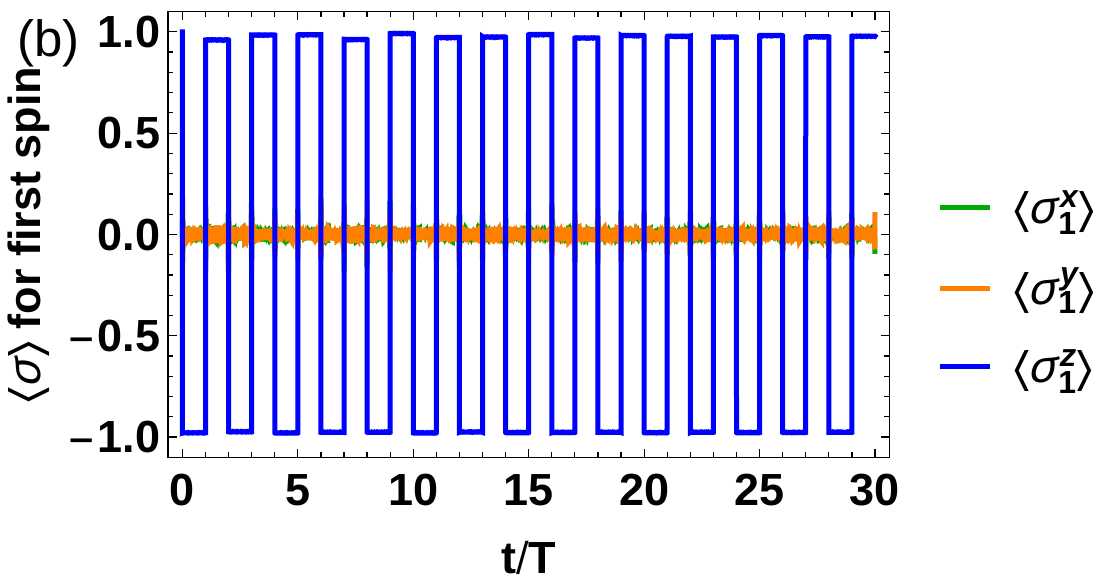}
\caption{Time evolution of spin vector components of first spin in $N=4$ spin chain for initial state $\ket{\uparrow\downarrow\uparrow\downarrow}$ with (a) Ising interactions with parameters $h^zT=0.05$, $\delta h^zT=0.05$, $\delta JT=0$, and (b) Heisenberg interactions with 64 H2I pulses with parameters $h^zT=2\times10^4$, $\delta h^zT=50$, $\delta JT=0$. In both panels, the exchange coupling is $JT=0.6$, and the pulse error is $\epsilon=0.1$.}\label{fig:DTTSB}
\end{figure}

\begin{figure*}
\includegraphics[width=0.4\columnwidth]{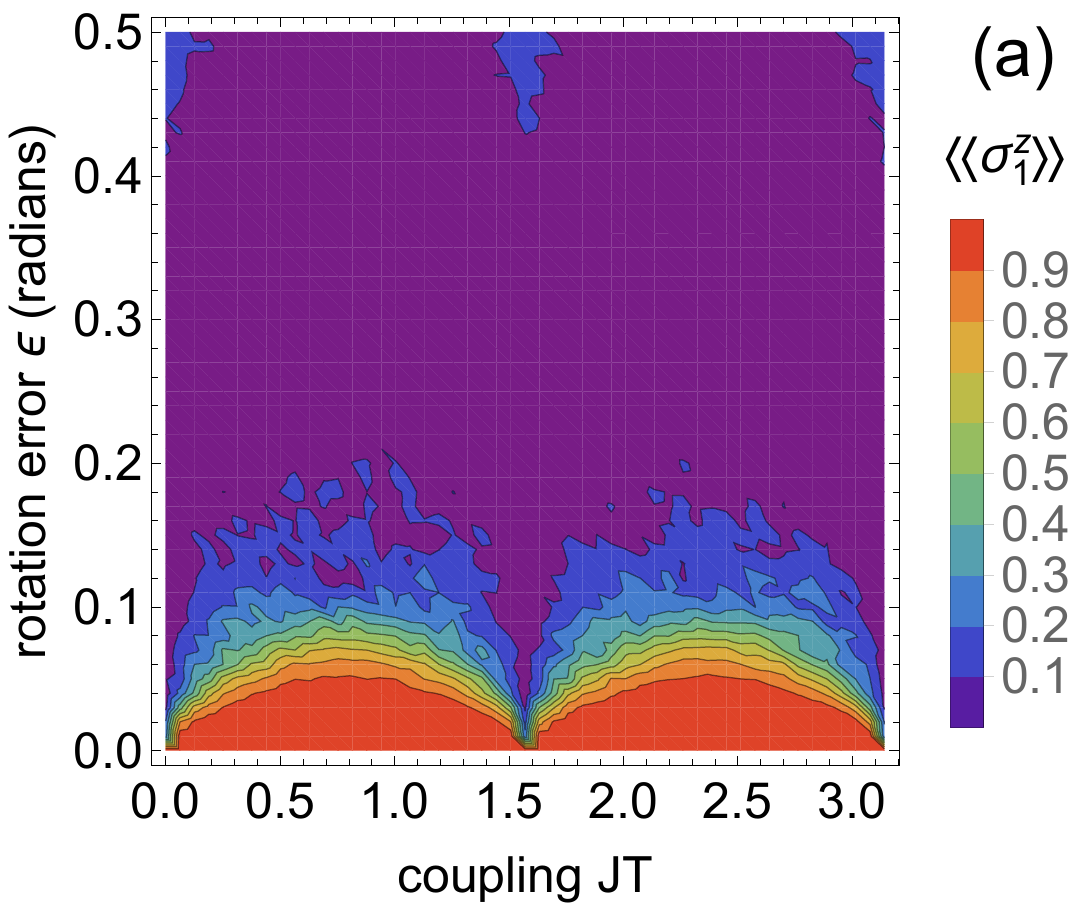}
\includegraphics[width=0.4\columnwidth]{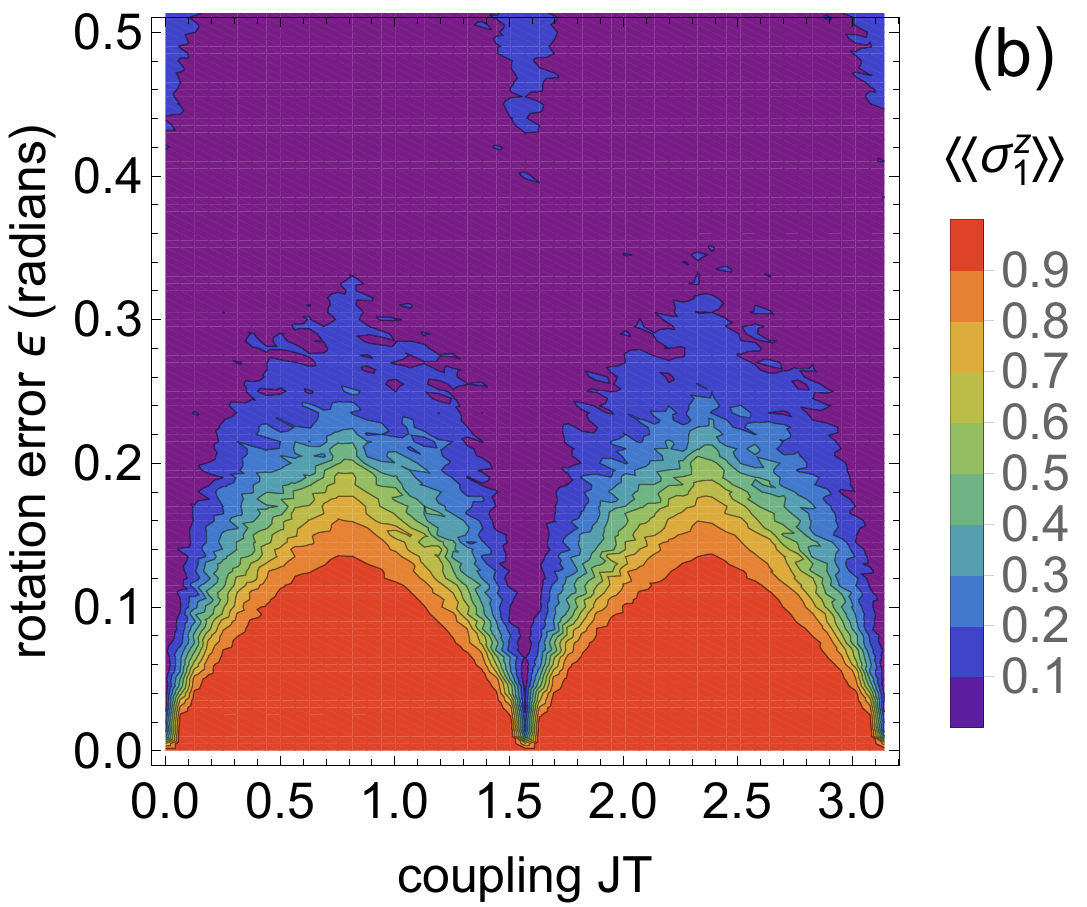}
\includegraphics[width=0.4\columnwidth]{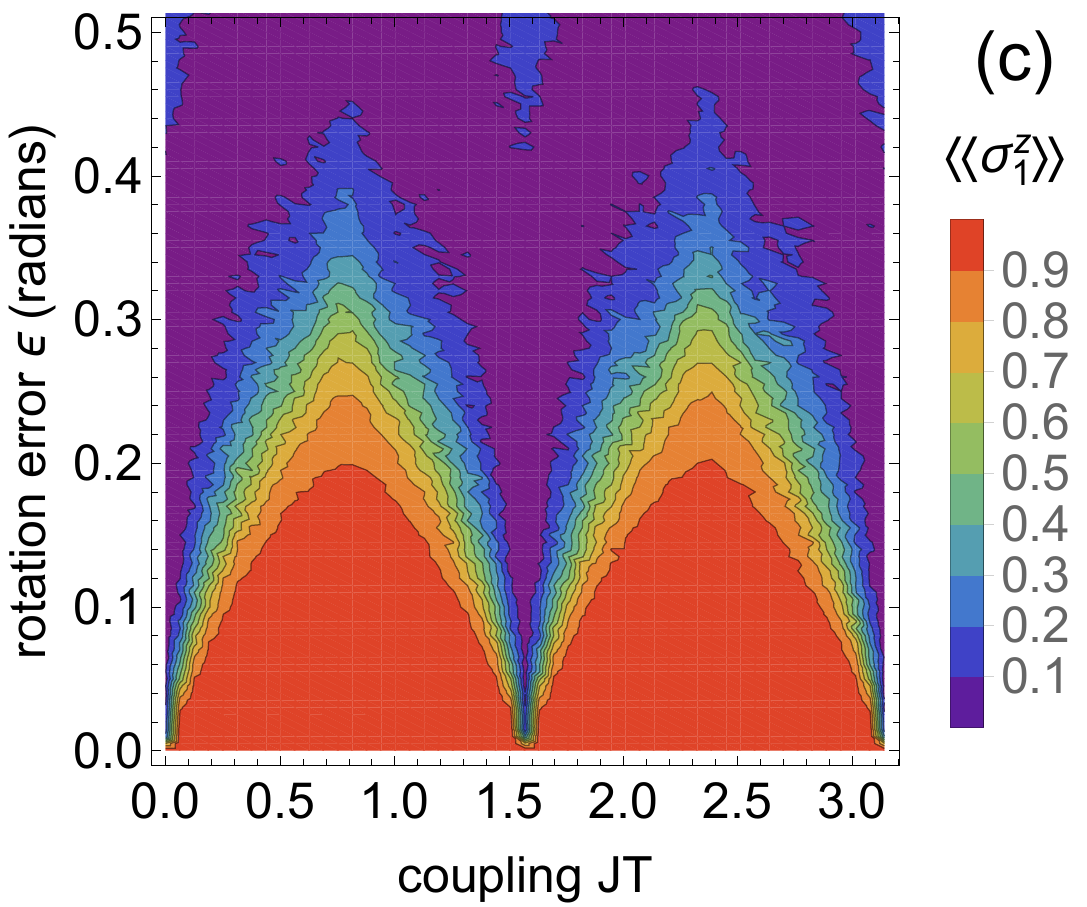}
\includegraphics[width=0.4\columnwidth]{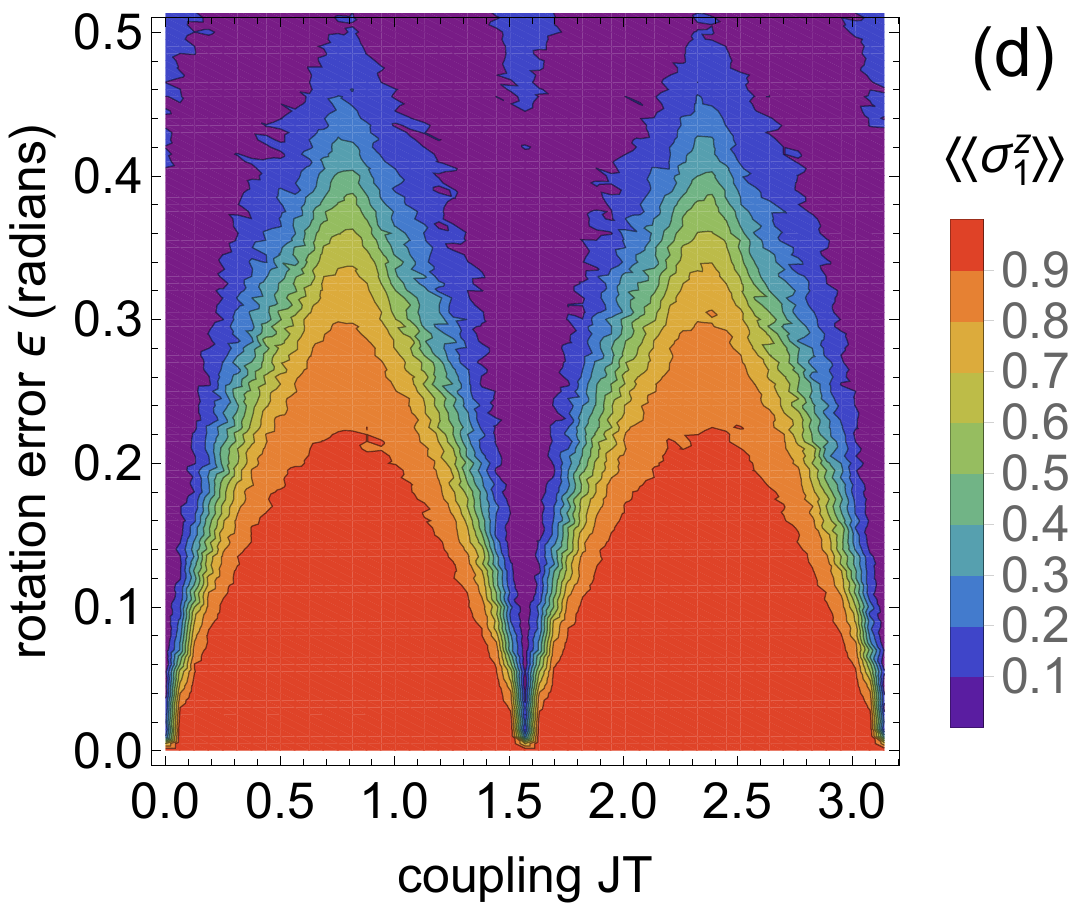}
\includegraphics[width=0.4\columnwidth]{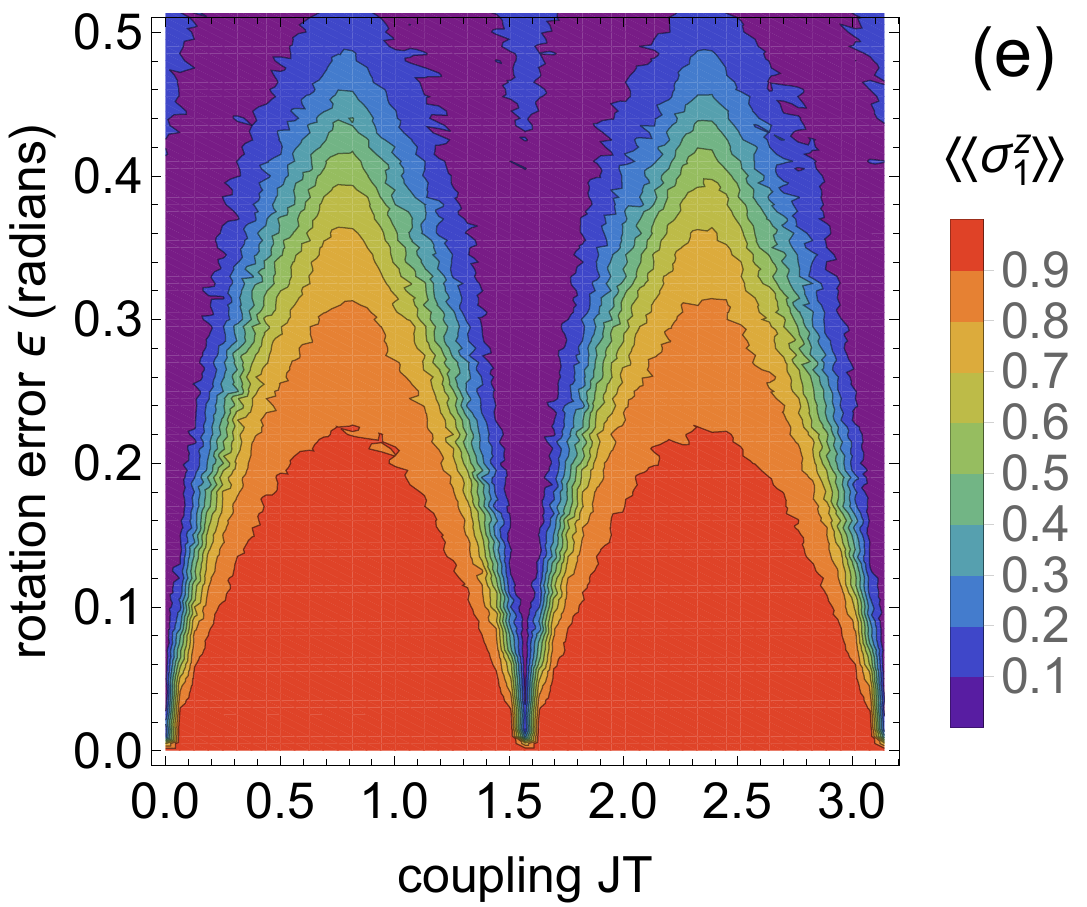}
\caption{Dependence on number of spins. Phase diagrams for Heisenberg spin chains driven by one Floquet pulse about $x$ and 128 H2I pulses about $z$ per period. The initial states are (a) $\ket{\uparrow\downarrow}$, (b) $\ket{\uparrow\downarrow\uparrow}$, (c) $\ket{\uparrow\downarrow\uparrow\downarrow}$, (d) $\ket{\uparrow\downarrow\uparrow\downarrow\uparrow}$, (e) $\ket{\uparrow\downarrow\uparrow\downarrow\uparrow\downarrow}$, and the system parameters are $h^zT=2\times10^4$, $\delta h^zT=50$, $\delta JT=0$.}\label{fig:heisenberg_changingnumspins}
\end{figure*}

\begin{figure*}
\includegraphics[width=0.45\columnwidth]{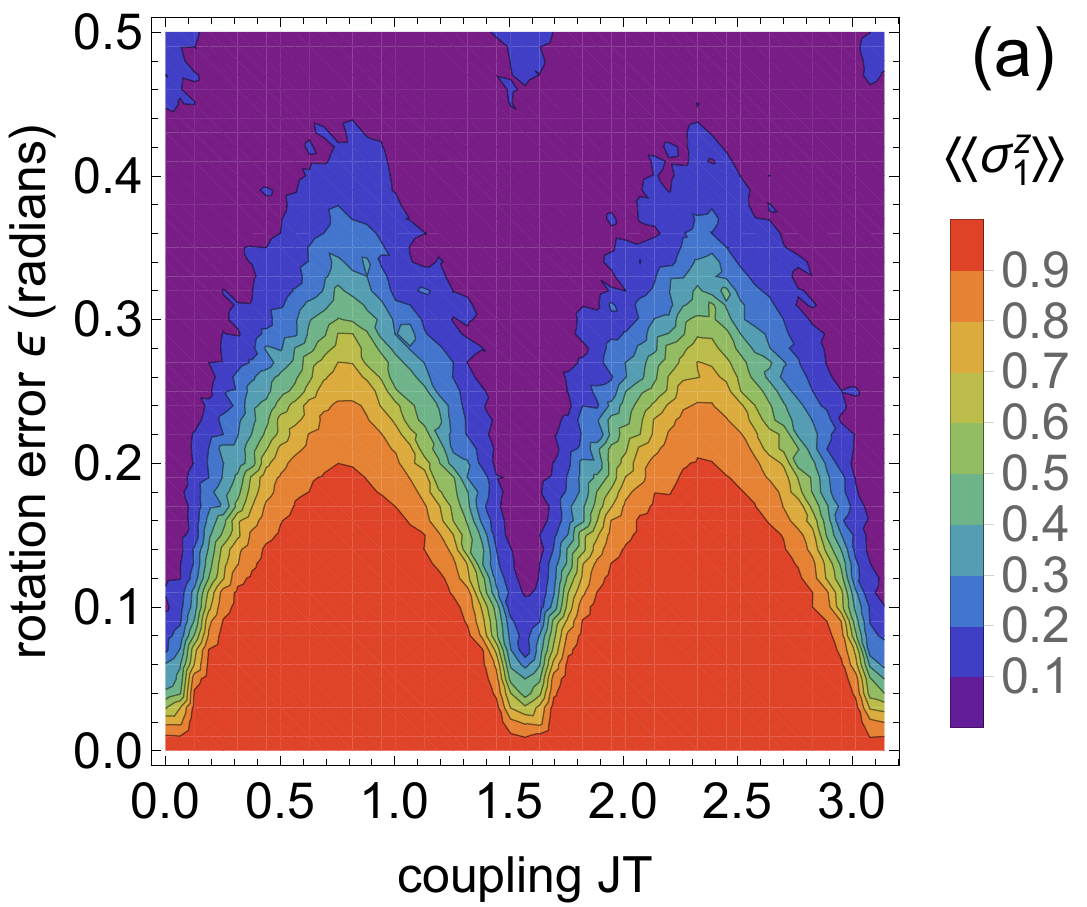}
\includegraphics[width=0.45\columnwidth]{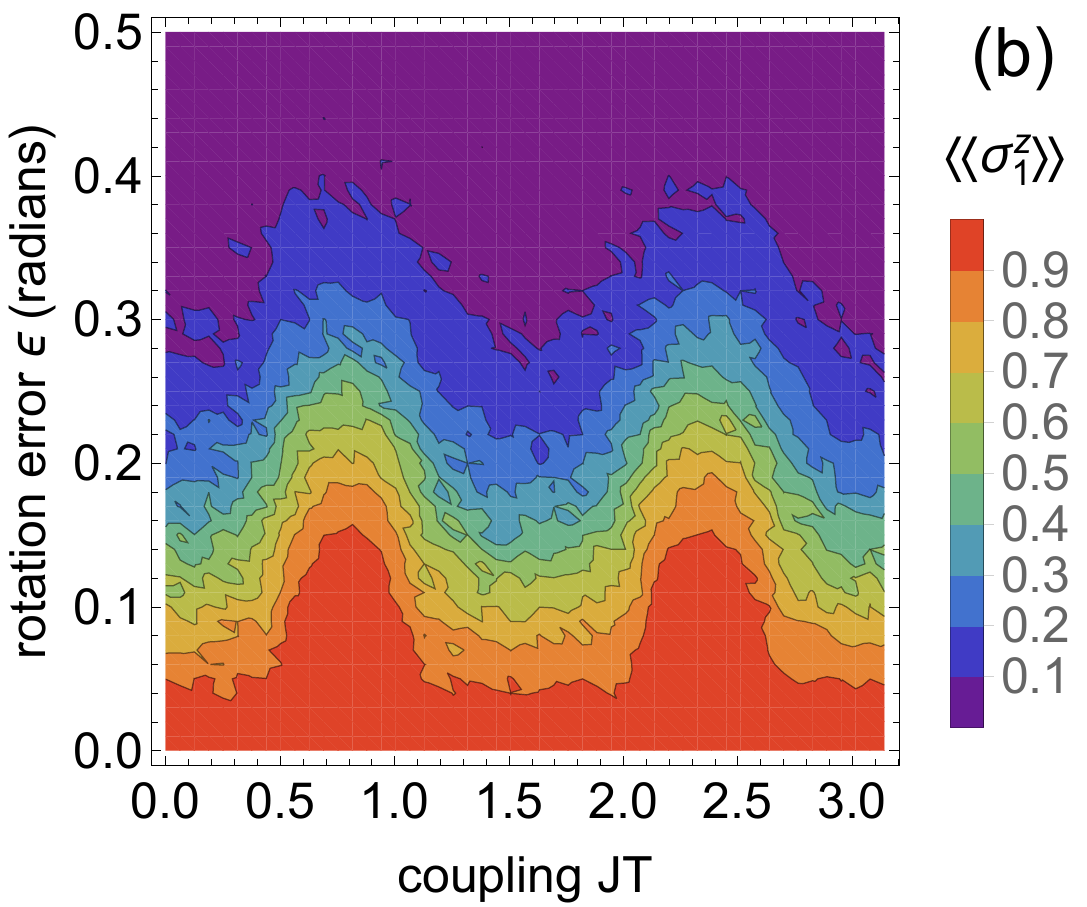}
\includegraphics[width=0.45\columnwidth]{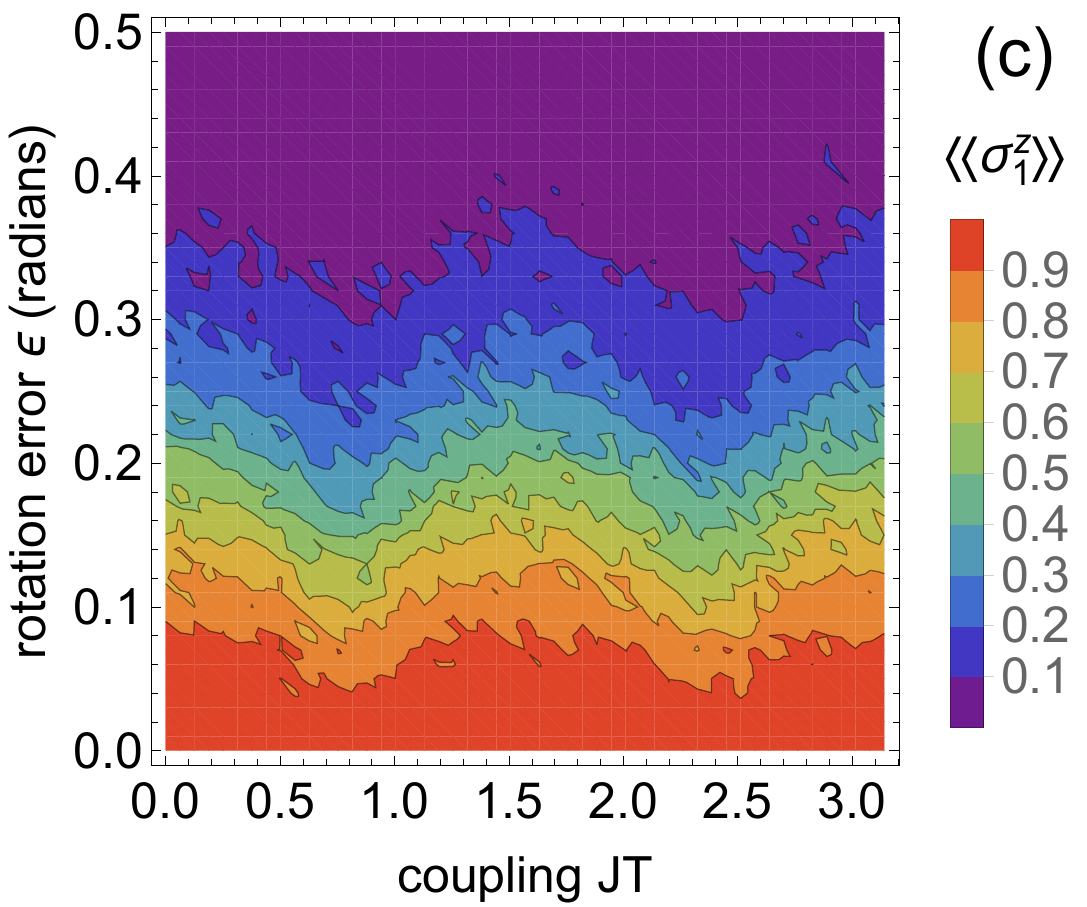}
\includegraphics[width=0.45\columnwidth]{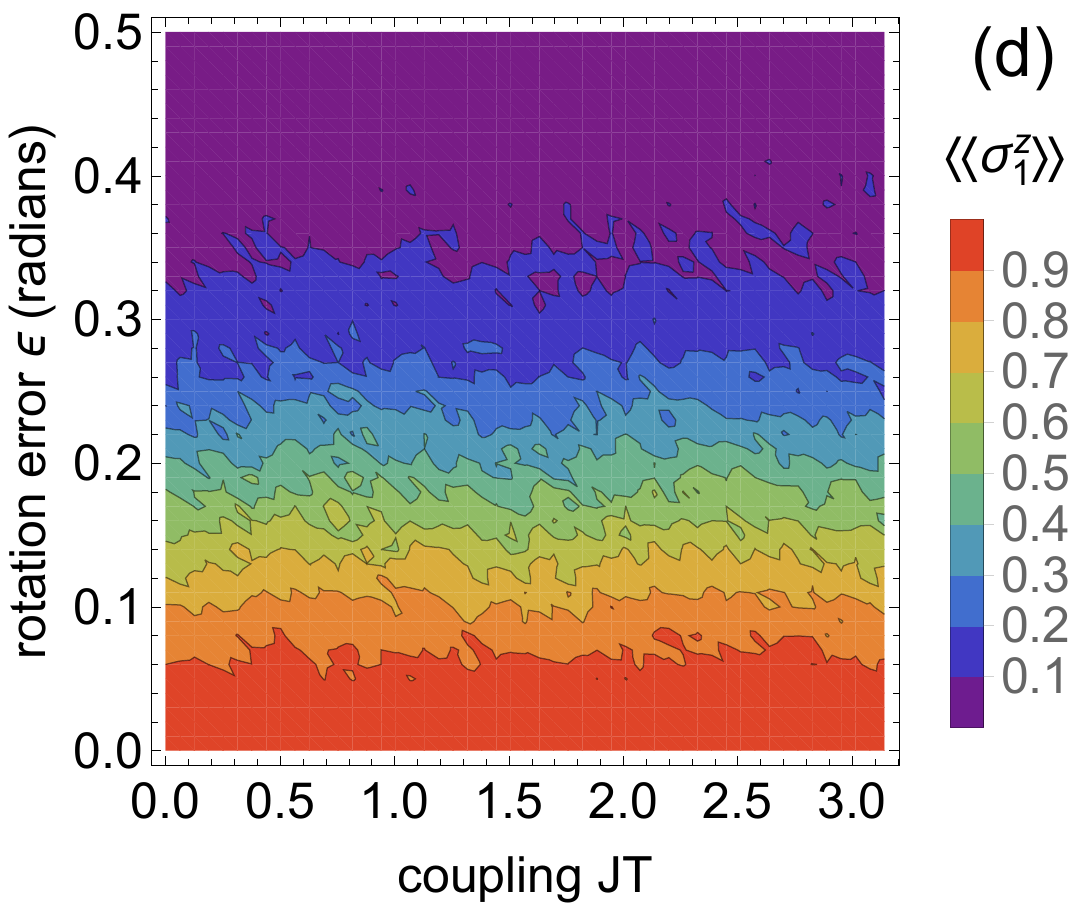}
\caption{Dependence on charge noise. Phase diagrams for a Heisenberg spin chain prepared in the state $\ket{\uparrow\uparrow\uparrow\uparrow}$ and driven by one Floquet pulse about $x$ and 128 H2I pulses about $z$ per period. The parameters are $h^zT=2\times10^4$, $\delta h^zT=50$, and (a) $\delta JT=0.1$, (b) $\delta JT=0.5$, (c) $\delta JT=1$, (d) $\delta JT=5$.}\label{fig:chargenoise}
\end{figure*}

\section{Conclusions}\label{sec:conclusions}

In conclusion, we have shown that discrete time crystals can be realized in periodically driven, exchange-coupled GaAs quantum dot arrays containing as few as three or four spins for typical levels of charge noise and nuclear spin noise. These phases arise if additional pulses are included during each drive period such that the Heisenberg interaction is effectively converted into an Ising interaction. We investigated the robustness of these phases over a wide range of pulse errors and coupling strengths. We demonstrated that by changing the rotation axes of the additional pulses and the direction of the external magnetic field, we can select which multi-spin states are preserved by the time crystal. Moreover, we can perform coherent spin rotations simply by altering the phase-stabilizing pulses after applying rotation pulses to all spins in the array. Our findings suggest that time crystals may be a useful tool in stabilizing and manipulating multi-spin states in quantum dot arrays.

\section*{Acknowledgments}

This work is supported by DARPA grant no. D18AC00025.

\appendix

\section{Discrete time translation symmetry breaking}\label{app:DTTSB}

In this Appendix, we confirm that the discrete $T$-periodic time translation symmetry of the driven spin chain is broken in the time crystal phase. This is demonstrated in Fig.~\ref{fig:DTTSB} for both the pure Ising spin chain case considered in Sec.~\ref{sec:ising} and the Heisenberg spin chain with H2I pulses and GaAs quantum dot parameters considered in Sec.~\ref{sec:heisenberg}. Fig.~\ref{fig:DTTSB}(a) is essentially the same as Fig.~\ref{fig:pointplots_ising}(a) in the main text, but here we show the full, continuous time evolution of the spin vector instead of only the spin vector values at the end of every second Floquet period. As in Ref.~\cite{Else_PRL16}, the fact that $\mean{\sigma_1^z}$ changes sign from one period to the next shows that the $T$-periodicity is broken to a $2T$-periodicity as is characteristic of a time crystal. The corresponding plot for the Heisenberg chain case, Fig.~\ref{fig:DTTSB}(b), exhibits the same symmetry breaking process when H2I pulses are used to convert the Heisenberg interaction into an effective Ising interaction. Fig.~\ref{fig:DTTSB}(b) corresponds to the $\epsilon=0.1$, $JT=0.6$ point of Fig.~\ref{fig:addingH2Ipulses2}(d).

\section{Dependence on number of spins and charge noise}\label{app:isingNdependence}

In Fig.~\ref{fig:heisenberg_changingnumspins}, we show how the phase diagram for a Heisenberg spin chain changes as the number of spins varies from $N=2$ to $N=6$. In each case, we consider a N\'eel-ordered initial state, and we switch off the coupling disorder. We see that as the number of spins is increased, the time crystal phase regions expand to include a larger range of pulse rotation errors. However, the basic qualitative features of the diagram remain unchanged.

The dependence of the phase diagram on charge noise (coupling disorder) is considered in Fig.~\ref{fig:chargenoise}. For levels of charge noise below the 1\% level (panel (a)), there is essentially no effect on the phase diagram, as can be seen by comparing to Fig.~\ref{fig:addingH2Ipulses2}(e). Since experimental levels of charge noise are in this regime, especially when barrier control is used \cite{Martins_PRL16}, this implies that $\delta J$ can be neglected. However, it is still interesting to consider what happens as the charge noise is increased beyond the experimentally relevant regime. This is shown in Fig.~\ref{fig:chargenoise}(b-d), where it is apparent that the time crystal phase becomes uniformly distributed across all values of the coupling $J$. This would of course be expected to occur when $\delta J\gg J$, but here we see that this is already the case for $\delta J\lesssim J$. It is also interesting to note that the range of $\epsilon$ over which this phase arises appears to converge to a constant value (in this case $\approx0.07$) as $\delta J$ becomes large.


\end{document}